\documentclass[twocolumn]{bmcart}

\RequirePackage{natbib}
\usepackage[utf8]{inputenc} 
\usepackage{endnotes}

\usepackage[british]{babel}
\def\includegraphics{}
\usepackage{graphicx}

\startlocaldefs
\endlocaldefs

\begin{document}

\begin{frontmatter}

\begin{fmbox}
\dochead{Research}


\title{On the Reliability of N-body Simulations}


\author[
   addressref={aff1},                   
   email={boekholt@strw.leidenuniv.nl, spz@strw.leidenuniv.nl}   
]{\inits{T}\fnm{T.} \snm{Boekholt}}
\author[
   addressref={aff1},
   email={spz@strw.leidenuniv.nl}
]{\inits{SPZ}\fnm{S.} \snm{Portegies Zwart}}


\address[id=aff1]{
  \orgname{Leiden Observatory, Leiden University}, 
  \street{PO Box 9513},                     %
  \postcode{2300 RA}                                
  \city{Leiden},                              
  \cny{The Netherlands}                                    
}


\begin{artnotes}
\end{artnotes}



\begin{abstractbox}

\begin{abstract} 

The general consensus in the \mbox{N-body} community is that statistical results of an ensemble of collisional \mbox{N-body} simulations are accurate, even though individual simulations are not. 
A way to test this hypothesis is to make a direct comparison of an ensemble of solutions obtained by conventional methods with an ensemble of true solutions. 
In order to make this possible, we wrote an \mbox{N-body} code called \texttt{Brutus}, that uses arbitrary-precision arithmetic. In combination with the Bulirsch--Stoer method, \texttt{Brutus} is able to obtain converged solutions, which are true up to a specified number of digits. \\
We perform simulations of democratic 3-body systems, where after a sequence of resonances and ejections, a final configuration is reached consisting of a permanent binary and an escaping star. We do this with conventional double-precision methods, and with \texttt{Brutus}; both have the same set of initial conditions and initial realisations. 
The ensemble of solutions from the conventional simulations is compared directly to that of the converged simulations, both as an ensemble and on an individual basis to determine the distribution of the errors. \\
We find that on average at least half of the conventional simulations diverge from the converged solution, such that the two solutions are microscopically incomparable.  
For the solutions which have not diverged significantly, we observe that if the integrator has a  bias in energy and angular momentum, this propagates to a bias in the statistical properties of the binaries.  
In the case when the conventional solution has diverged onto an entirely different trajectory in phase-space, we find that the errors are centred around zero and symmetric; the error due to divergence is unbiased, as long as the time-step parameter, $\eta \le 2^{-5}$ and when simulations which violate energy conservation by more than 10$\%$ are excluded. 
For resonant 3-body interactions, we conclude that the statistical results of an ensemble of conventional solutions are indeed accurate.

\end{abstract}




\end{abstractbox}
\end{fmbox}

\end{frontmatter}


\section{Introduction}

Analytical solutions to the \mbox{N-body} problem are known for $N=2$, which are the familiar
conic sections. 
Also, for several systems possessing symmetries, analytical solutions have been found,  
for example the equilateral triangle \citep{Lagrange1772}. 
For a more general initial configuration, solutions have to be obtained by means of numerical integration. 
Given an initial \mbox{N-body} realisation, one can calculate all mutual forces and subsequently the net acceleration of each particle. 
Different integration methods exist which take the accelerations, and update the positions and velocities to a time $t+\Delta t$, with $\Delta t$ the time-step size. 
This process is repeated until the end time is reached.

Miller \citep{1964ApJ...140..250M} recognised that obtaining the solution to an \mbox{N-body} problem by numerical integration is difficult. 
This is caused by exponential divergence. 
Consider a certain \mbox{N-body} problem, i.e. $N$ point-particles, each with a given mass, position and velocity. 
This system evolves with time in a definite and unique way. 
If one goes back to the initial state and slightly perturbs only one coordinate of a single particle, the perturbed \mbox{N-body} problem will also have a definite and unique but different solution than the original one. 
When the two solutions are compared as a function of time, it is observed that differences can grow
exponentially \citep{1964ApJ...140..250M, 1986LNP...267..212D, 1993ApJ...415..715G, 2002JSP...109.1017H}. 
If the initial perturbation is due to a numerical error, the calculated solution will also diverge away from the true solution.

Several authors have estimated the time-scale of this divergence \citep{1993ApJ...415..715G, 2002JSP...109.1017H}, and arrived at an e-folding time-scale of the order a dynamical, crossing time.
Simulation times of interest are typically much longer than a crossing time and therefore staying close
to the true solution is numerically challenging. 

If the result of a direct \mbox{N-body} simulation of for example a star cluster, has diverged away from the true solution, the result may well be meaningless \citep{1993ApJ...415..715G}. 
The general consensus however, is
that statistically the results are representative for the true
solution to the \mbox{N-body} problem \citep{1979A&A....76..192S, 1991pscn.proc...47H, 1993ApJ...415..715G}.
The underlying idea is that the statistics of an ensemble of \mbox{N-body} simulations 
are representative for the true statistics, obtained by an ensemble of true solutions, 
with the same set of initial conditions.
We regard this the hypothesis we want to test. 

One way to test this hypothesis is to directly compare statistics obtained by conventional methods, with the statistics obtained from an ensemble of true solutions. To obtain true solutions, we wrote an \mbox{N-body} code which can solve the \mbox{N-body} problem to arbitrary precision. 

Such a code can be realised if the different sources of error are controlled.
The error has contributions from the time discretisation
of the integrator and the round-off due to the limited precision of
the computer \citep{1979CeMec..20..209Z}. Another possible source of error is in the initial conditions, for example the configuration of the Solar System is only approximately known \citep{2002MNRAS.336..483I}. However, if the initial condition is a random realisation of a distribution function, this is less often a problem. 
Using the Bulirsch--Stoer method \citep{springerlink:10.1007/BF01386092, 1965SJNA....2..384G}, the discretisation error can be controlled to stay within a specified tolerance.  
Using arbitrary-precision arithmetic instead of conventional double-precision or single-precision, the round-off error can be reduced by increasing the number of digits.  

We obtain converged solutions to the \mbox{N-body} problem by decreasing the Bulirsch--Stoer tolerance and increasing the number of digits systematically. 
We define a converged solution in our experiments as a solution for which the first specified number of decimal
places of every phase-space coordinate in our final configuration in the \mbox{N-body} experiment becomes independent of the length of the mantissa and the Bulirsch--Stoer tolerance.  
We explain the method of convergence in Sec.~\ref{Sect:Methods} and we give examples of the procedure in Sec.~\ref{Sect:Validation}. 

Using this new, brute force \mbox{N-body} code which we call \texttt{Brutus}, we test the reliability of \mbox{N-body} simulations by a controlled numerical experiment which we describe in Sec.~\ref{Sect:Experiments}. 
In this experiment we perform a series of resonant 3-body simulations, where the term resonant implies a phase or multiple phases during the interaction where the stars are more or less equidistant \citep{1983ApJ...268..319H}. These phases are intermingled by ejections, where a binary and single star are clearly separated. 
We perform the simulations with conventional double-precision, and  with arbitrary-precision to reach the converged solution. 
In Sec.~\ref{Sect:Results}, the solutions are compared individually to investigate the distribution of the errors. We also compare the global statistical distributions using two-sample Kolmogorov--Smirnov tests \citep{1933Kolmogorov, 1948Smirnov}. 

\section{Methods}\label{Sect:Methods}

\subsection{The benchmark integrator}

The gravitational \mbox{N-body} problem aims to solve Newton's equations of motion under gravity for $N$ stars \citep{Newton:1687}. A popular integrator to perform this task is the fourth-order \texttt{Hermite} predictor-corrector scheme \citep{1992PASJ...44..141M}, using double-precision arithmetic.  
The experiments we discuss in Sec.~\ref{Sect:Experiments} will use this integrator as a benchmark test.
We adopt a shared, adaptive time-stepping scheme with the following criterion: 

\begin{equation}
  \Delta t = \eta \min{ \sqrt{ \Delta r_{ij}/\Delta a_{ij} } }.    
  \label{eq:1}
\end{equation}

\noindent Here $\eta$ is the time-step parameter and $\Delta r_{ij}$ and $\Delta a_{ij}$ are the relative distance and acceleration for the pair of particles $i$ and $j$. We implement no further constraints on the time-step size. 

To test how inaccurate we are allowed to integrate while still obtaining accurate statistics \citep{1979A&A....76..192S, 1992MNRAS.259..505Q} we vary the time-step parameter $\eta$, to obtain statistics from conventional simulations with different precision. 

\subsection{The \texttt{Brutus} N-body code}

The results obtained with the benchmark integrator are compared to those obtained with \texttt{Brutus}, which uses an arbitrary-precision library \endnote{We use the open-source library GMP: http://gmplib.org/}. With this library we can specify the number of bits, $L_{w}$, used to store the mantissa, while the exponent has a fixed word-length. The length of the mantissa can be specified and increased, with the aim of controlling the round-off error. 

The integration of the equations of motion is realised using the Verlet-leapfrog scheme \citep{PhysRev.159.98}. 
The time-step is shared among all particles, but varies for every step according to equation \ref{eq:1}.  

To control the discretisation error,  we implemented the Bulirsch--Stoer (BS) method, which uses iterative integration and polynomial extrapolation to infinitesimal time-step size \citep{springerlink:10.1007/BF01386092,1965SJNA....2..384G}. 
An integration step is accepted, when two subsequent BS iterations have converged to below the BS tolerance level, $\epsilon$. 

The time-step parameter $\eta$ and the BS tolerance $\epsilon$, both influence the performance. If $\eta$ is too big, convergence may not be achieved for any tolerance. If $\eta$ is too small, the many integration steps will render the integration too expensive. There is an optimal value for $\eta$ as a function of $\epsilon$. We measured this relation empirically, which results in: 

\begin{equation}
  \log_{10} \eta = A \log_{10} \epsilon + B.
    \label{eq:2}
\end{equation}

\noindent For $\epsilon < 10^{-50}$ the powerlaw converges to $A=0.029$ and $B=0.45$. Extrapolating this relation to $\epsilon > 10^{-50}$ will cause the time-step size to become larger than the time scale for the closest encounter in the system. Therefore this relation saturates to a flatter powerlaw for $\epsilon > 10^{-50}$ with $A=0.012$ and $B=-0.40$. 
Compared to a fixed value for $\eta$, this relation speeds up the iterative procedure by about a factor three or more. 
The code is implemented as a community code in the AMUSE framework \citep{2012ASPC..453..317P} under the name \texttt{Brutus}. 

\subsection{Method of convergence}\label{MoC}

\begin{figure*}[t]
\centering
\begin{tabular}{l}
\includegraphics[scale=0.77]{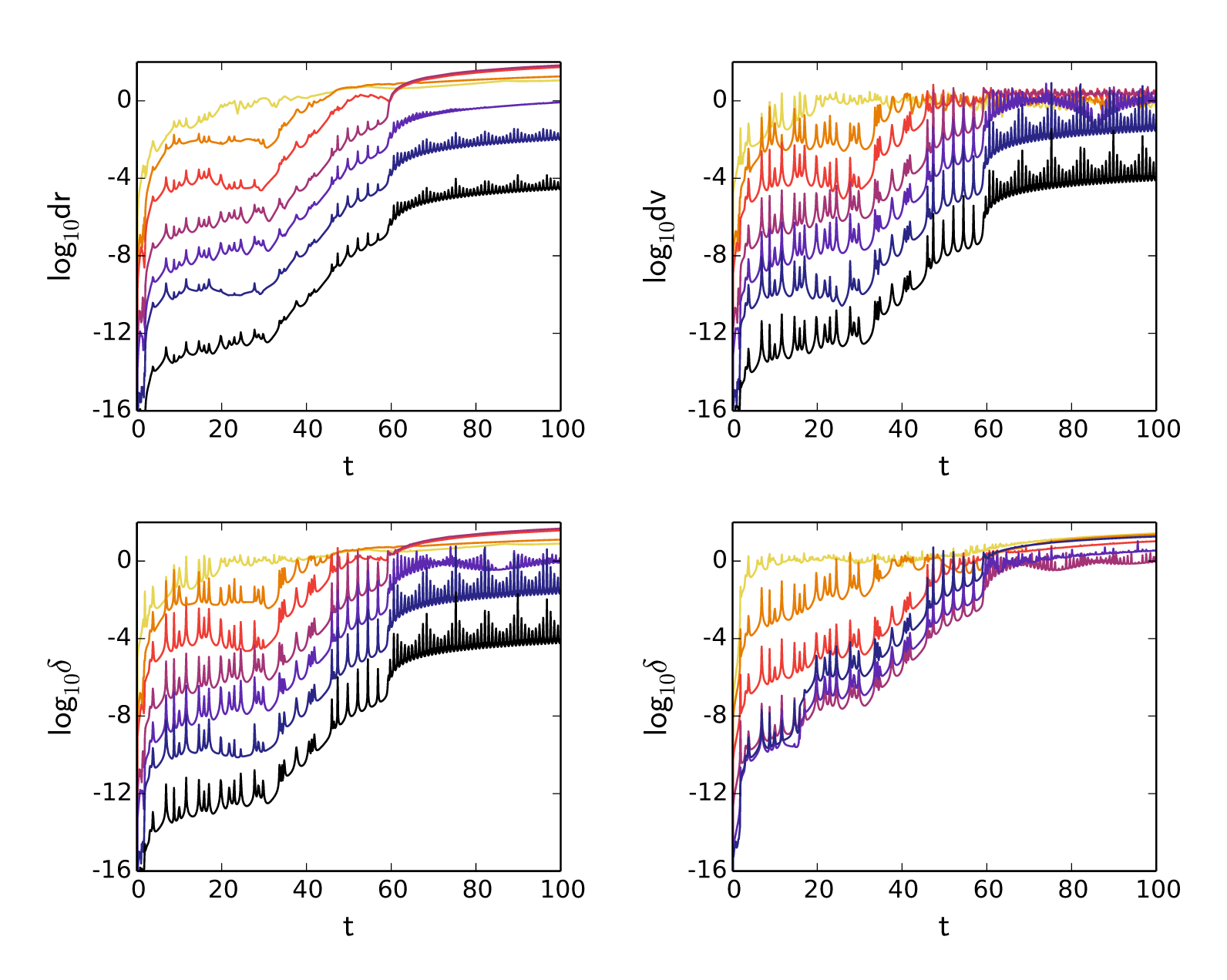} \\
\end{tabular}  
\caption{ \csentence{Exponential divergence in the Pythagorean problem.}
In the top two panels and the lower left panel, \texttt{Brutus} is compared with \texttt{Brutus} with increasing precision. The yellow curves (curves at the top) compare a tolerance of $10^{-2}$ with $10^{-4}$, the orange curves (second curve from the top) compare $10^{-4}$ with $10^{-6}$ and so on. The word-length is a function of the tolerance as in Eq.~\ref{eq:4}. In the top left panel  we show the distance in position-space, in the top right panel in velocity-space and in the bottom left panel in the full phase-space (all normalized by the number of stars and coordinates). The lower right panel compares the converged solution (black and lowest curve in the other plots), with \texttt{Hermite} solutions with time-step parameters $\eta = 2^{-3}, 2^{-5}, 2^{-7}$ up to $2^{-13}$, with a color sequence similar as in the other panels. }
\label{fig:pyth}
\end{figure*}

For every simulation we have to define the BS tolerance parameter $\epsilon$ and the word-length $L_{w}$. 
In an iterative procedure we vary both parameters
systematically, each time carrying out a simulation until $t=t_{\rm end}$. 
We subsequently calculate the phase space distance, $\delta^2_{A,B}$, between two solutions A and B:

\begin{equation}
  \delta^{2}_{A,B}={1 \over 6N} \sum\limits_{i=1}^N \sum\limits_{j=1}^6 \left( q_{A,i,j} - q_{B,i,j} \right)^{2}.
    \label{eq:3}
\end{equation}

\noindent The first summation is over all particles and the second summation is over the six phase-space coordinates denoted by $q$ \citep{1964ApJ...140..250M}. We normalise by $6N$, so that $\delta$ represents the average difference per phase-space coordinate between two solutions A and B. In our experiments we adopt Hénon units \endnote{Formerly known as \mbox{N-body} units. Introduced by D. Heggie at MODEST14.} \citep{1971Ap&SS..14..151H, 1986LNP...267..233H}, in which the typical values for the distance and velocity are of the same order. We will also use the distance in just position or just velocity space as they might behave differently.  

We consider the solutions A and B to be converged when $\delta_{A,B}<10^{-p}$ at all times during the simulation. Note that converged in this case means convergence of the total solution, contrary to convergence per integration step as in the previous section. 
This criterion for convergence is roughly equivalent to comparing the first $p$ decimal places of the positions and
velocities for all $N$ stars, in two subsequent calculations A, B.
In most of our experiments we adopt $p=3$, i.e. all coordinates have to converge to about three decimal places or more. We perform a subset of simulations with $p=15$ to investigate the effect of small errors (see Sec.~\ref{symresult}). 

Each simulation starts by specifying the initial positions and
velocities of $N$ stars in double-precision (see Sec.~\ref{Sect:Experiments}). 
The simulation is carried out with the parameter set ($\epsilon$, $L_{w}$).   
We start each simulation with $\epsilon=10^{-6} $ and $L_w=56$\,bits. This corresponds to 
a level of accuracy similar to what we reach
with the conventional \texttt{Hermite} integrator.
After this simulation, we increase $L_w$, for example to 72 \,bits ($\sim$ 22 decimal places), redo the
simulation and calculate $\delta^2$.  We repeat this procedure until
$\delta < 10^{-p}$.  When this is achieved, we have obtained a
solution in which the round-off error is below a specified number of decimal places for
this particular value of $\epsilon$.

We now reduce the tolerance parameter $\epsilon$, for example by a factor of 100, and repeat the procedure of increasing $L_w$.  
This series will again lead to a converged solution, but this time it is obtained using a smaller $\epsilon$, and is likely to
be different than the previous converged solution. 
We continue decreasing the value of $\epsilon$ by
factors of 100 and repeat the procedure, until two subsequent
iterations in $\epsilon$ lead to a converged solution with a value of $\delta < 10^{-p}$. 
By this time we are assured of having a solution to the gravitational \mbox{N-body} problem, 
that is accurate up to at least $p$ decimal places.

In practice, we speed up the procedure by writing the word-length as a function of BS tolerance. Consider for example a BS tolerance of $10^{-20}$. 
To reach convergence up to this level, we need at least 20 decimal places. Adding an extra buffer of 10 digits gives a total of 30 digits, 
or equivalently a word-length of about 112\,bits. For this example, 112\,bits turns out to be a good minimum word-length. For a first estimate of the word-length, we use: 

\begin{equation}
  L_w=4 \left| {\log_{10} \epsilon} \right| + 32 \: \mathrm{bits}.
\label{eq:4}
\end{equation}

\noindent With this relation, we will only have to specify a single parameter $\epsilon$, which directly controls the discretisation error and indirectly controls the round-off error. 
For most of the systems in our experiment the discretisation error turns out to be the dominant source of error and as a consequence $\epsilon$ has to decrease quite drastically. 
When $\epsilon$ decreases, $L_w$ increases, even up to the point that there are many more digits available than really needed to control the round-off error. 
In the case when the discretisation error dominates, the above defined minimum word-length for a given BS tolerance will result in the converged solution. When the round-off error dominates the word-length should be varied independently. 

\section{Validation and performance}\label{Sect:Validation}

\subsection{The Pythagorean problem}

\begin{figure*}[t]
\centering
\begin{tabular}{l}
\includegraphics[scale=0.88]{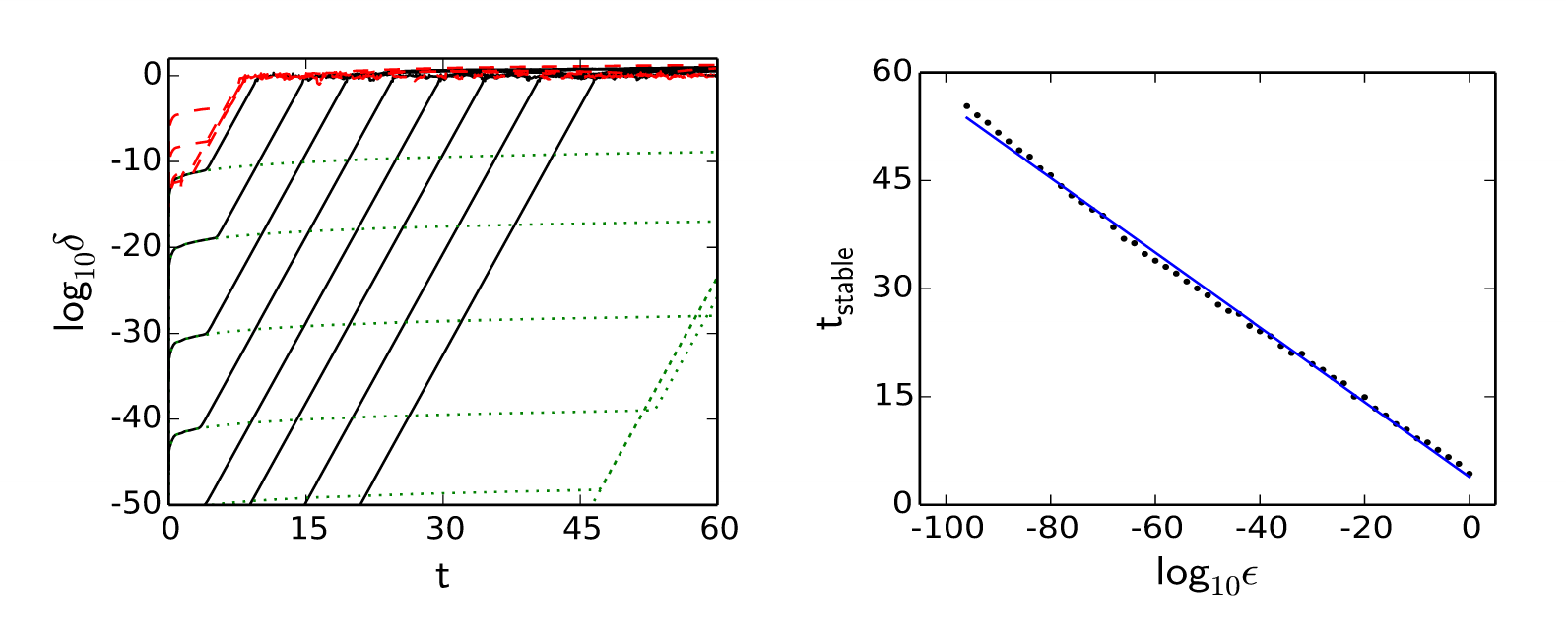} \\
\end{tabular}  
\caption{\csentence{Divergence for the equilateral triangle configuration.}
In the left panel we show the divergence as a function of time. The solid, black curves compare \texttt{Brutus} solutions with increasing precision, where subsequent precisions are increased by 10 orders of magnitude and where the word-length is a function of tolerance as in Eq.~\ref{eq:4}. The dotted, green curves show results for similar simulations, but with a much longer, fixed word-length of 512\,bits. The initial power law phase of divergence lasts longer in this case. The exponential divergence becomes dominant when the round-off error has had time to accumulate to become of the order the discretisation error. The dashed, red curves compare the highest precision \texttt{Brutus} solution with \texttt{Hermite} solutions with time-step parameters 10$^{-1}$, 10$^{-2}$, 10$^{-3}$ and 10$^{-4}$. In the right panel we show for \texttt{Brutus}, the duration for which the triangular configuration remains intact as a function of Bulirsch--Stoer tolerance $\epsilon$. Note that the time is in units of the period of one complete rotation of the system. The small scatter in the data is due to the discrete times at which we check the triangular configuration. }
\label{fig:triangle}
\end{figure*}

To show that our method works, we adopt the Pythagorean 3-body system \citep{1913AN....195..113B}. 
Previous numerical studies have shown that this system dissolves into a binary and an escaper \citep{1967AJ.....72..876S, 1994CeMDA..58....1A}.
After many complex, close encounters the dissolution happens at about $t=60$ time units \citep{1986LNP...267..212D}, or about $16$ crossing times.  

We adopt the initial conditions for the Pythagorean problem and integrate up to $t=100$. 
To illustrate how the method works we start with a high tolerance and short word-length, ($\epsilon=10^{-2}$, $L_{w}=40$\,bits), which is less precise than double-precision.
In Fig.~\ref{fig:pyth}, this calculation is compared to a simulation with ($\epsilon=10^{-4}$, $L_{w}=48$\,bits), through the yellow (upper) curves in the first three panels. 
After the first BS integration step, $\delta$ obtains a value of the order of the BS tolerance, and continues to increase due to exponential divergence, to eventually exceed $\delta$ $\sim$ $10^{-1}$, after which the errors become on the order of the typical distance and speed in the system. 

In the following step, we repeat the calculation with a precision of ($\epsilon=10^{-6}$, $L_{w}=56$\,bits), and compare the result with the calculation using ($\epsilon=10^{-4}$, $L_{w}=48$\,bits). The comparison is represented by the orange curves (second from above) in Fig.~\ref{fig:pyth}. The overall behaviour of $\delta$ is similar, but the system diverges at a later time due to a higher initial precision. 

We continue the iterative procedure until a converged solution has been obtained. In the first three panels of Fig.~\ref{fig:pyth}, it can be seen that subsequent simulations with higher precision shift the curve to lower values of $\delta$. Superposed on the steady growth of the error are sharp spikes, where the error grows by several orders of magnitude, after which the error restores again \citep{1964ApJ...140..250M}. These spikes are dominated by errors in the velocity, as can be deduced by comparing the magnitude of the spikes in position and velocity-space. Eccentric binaries which are out of phase when comparing two solutions cause large, periodic errors in the velocity. We finish the procedure when a solution is obtained for which the criterion for convergence is fulfilled, considering the magnitude of the error between the sharp spikes (bottom, black curves). 

In the bottom right panel of Fig.~\ref{fig:pyth}, we compare solutions obtained by the \texttt{Hermite} integrator to the converged solution. The different curves belong to different time-step parameters; $\eta=2^{-3}$, $2^{-5}$, $2^{-7}$ up to $2^{-13}$. Note that for a time-step parameter $\eta < 2^{-9}$, the curve is not shifted to lower values of $\delta$, but even increases again. At this point round-off error becomes important, making the solution less accurate. The final close encounter in the Pythagorean problem occurs around 60 time units, after which a permanent binary and an escaper are formed. The \texttt{Hermite} integrator is able to accurately reproduce the evolution up to this point, but not subsequently, because $\delta$ has increased to values of order unity or higher. This can be explained by a small error in the final close encounter between all three stars, such that the direction of the escaper is slightly different. 

To obtain the converged solution up to the first three decimal places, a tolerance of $10^{-14}$ and a word-length of 88\,bits were needed. The simulation was about twice as slow compared to the \texttt{Hermite} simulation with $\eta=2^{-9}$. The \texttt{Hermite} simulation, however, had a slightly different solution and a final, relative energy conservation of $10^{-8}$, Decreasing the value of $\eta$ will improve the level of energy conservation, but due to round-off error $\delta$ will not decrease. 

\subsection{The equilateral triangle}

\begin{figure*}[t]
\centering
\begin{tabular}{l}
\includegraphics[scale=0.88]{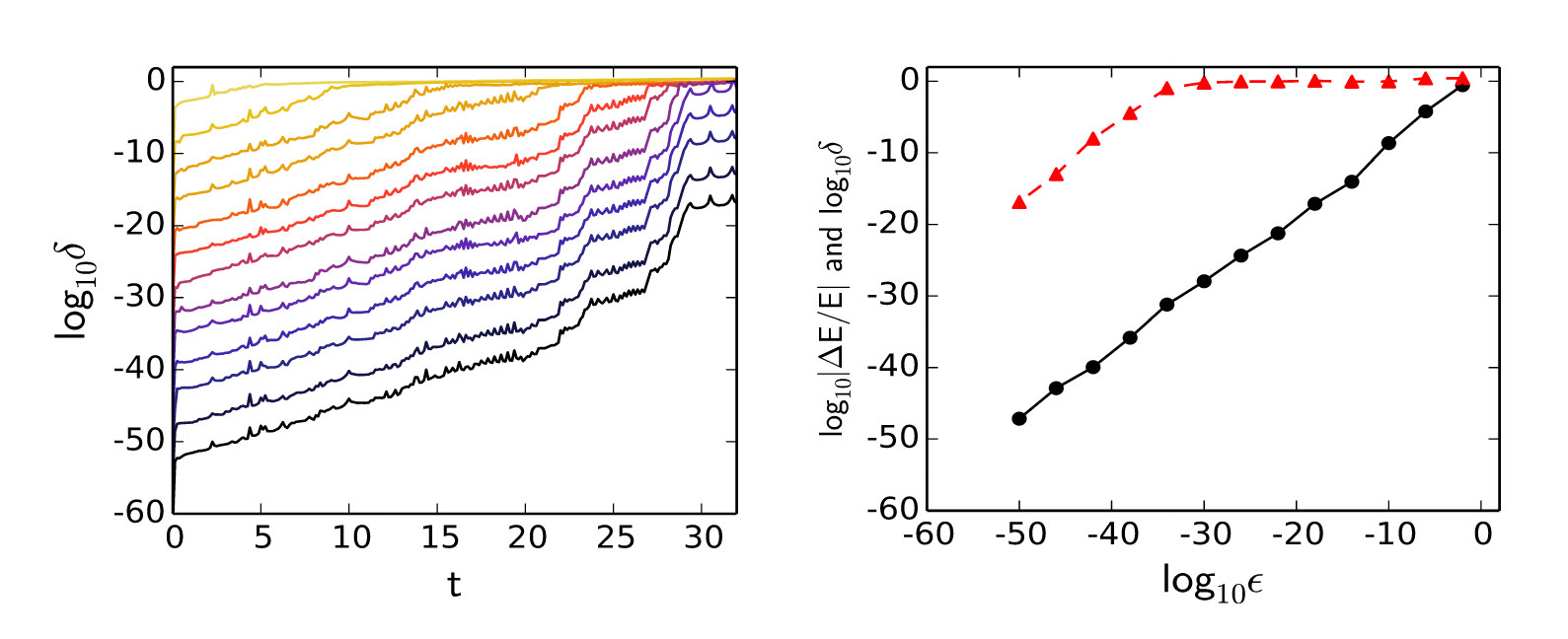} \\
\end{tabular}  
\caption{\csentence{Exponential divergence in a 16-body cluster.}
In the left panel we illustrate the exponential divergence between \texttt{Brutus} simulations with increasing precision. In the right panel we show the final relative energy conservation (black bullets, solid line) and the final normalized phase space distance between two subsequent simulations (red triangles, dashed line) versus the Bulirsch--Stoer tolerance parameter $\epsilon$. The solution starts to converge at a level of final relative energy conservation of $\sim 10^{-34}$.   }
\label{fig:plummer}
\end{figure*}

As a second test case, we adopt the 3-body equilateral triangle as an initial condition \citep{Lagrange1772}. In the exact solution this configuration remains intact, but small perturbations, such as produced by numerical errors,  quickly cause the triangle to fall apart. For this problem, we also have a source of error in the initial conditions. Whereas the Pythagorean problem can be set up using integers, the initial condition for the equilateral triangle contains irrational numbers. To control the error in the initial condition, we calculate the initial coordinates with the same word-length as used for the simulation.

In the left panel of Fig.~\ref{fig:triangle}, a similar diagram is shown as for the Pythagorean problem in the lower left panel of Fig.~\ref{fig:pyth}. The starting precision is $\epsilon=10^{-10}$ and the word-length is a function of $\epsilon$ as in Eq.~\ref{eq:4}. Subsequent simulations are performed with a 10 orders of magnitude higher precision. For a short initial phase of 5 time units, the rate of divergence follows a power law. At later time, the solutions start to diverge exponentially with a characteristic rate independent of the tolerance and word-length. To investigate this transition, we redo the simulations with a large, fixed word-length of 512\,bits (green dotted curves). This way, we reduce the amount of round-off error. As a consequence the rate of divergence is first dominated by the accumulation of discretisation errors and this phase lasts for a longer time, until the transition in the behaviour of the divergence, is reached, but now  at $\sim 45$ time units. 
The time of the transition depends on word-length. Why the exponential divergence starts once the round-off error has kicked in, is a question that is still under investigation. 

The red dashed curves in the same diagram in Fig.~\ref{fig:triangle} give the results of the fourth-order \texttt{Hermite}, which are compared with the most precise \texttt{Brutus} simulation (with $\epsilon=10^{-80}$, $L_{w}=352$\,bits). The time-step parameter $\eta=$ $10^{-1}$, $10^{-2}$, $10^{-3}$ and $10^{-4}$ for subsequent curves. The \texttt{Hermite} integrations show a similar behaviour as the \texttt{Brutus} results, which could imply that the rate of  divergence is a physical property of the configuration, rather than a property of the integrator. 

In the right panel of Fig.~\ref{fig:triangle} we show the duration for which the triangular configuration remains intact as a function of BS tolerance. For this experiment we halt the simulation when the distance between any two particles has increased or decreased by 10$\%$, after which the triangle falls apart quickly. This diagram also illustrates the linear relation between accuracy and time in this system, which is caused by the constant number of digits being lost during every unit of time. The small scatter is due to the discrete times at which we check the triangular configuration. The solid, blue line is a fit to the data and its slope is $-0.52(3)$, which is equivalent to a loss of $1.9(1)$ digits per cycle. 

\subsection{A Plummer distribution with N=16}

\begin{figure*}[t]
\centering
\begin{tabular}{l}
\includegraphics[scale=0.88]{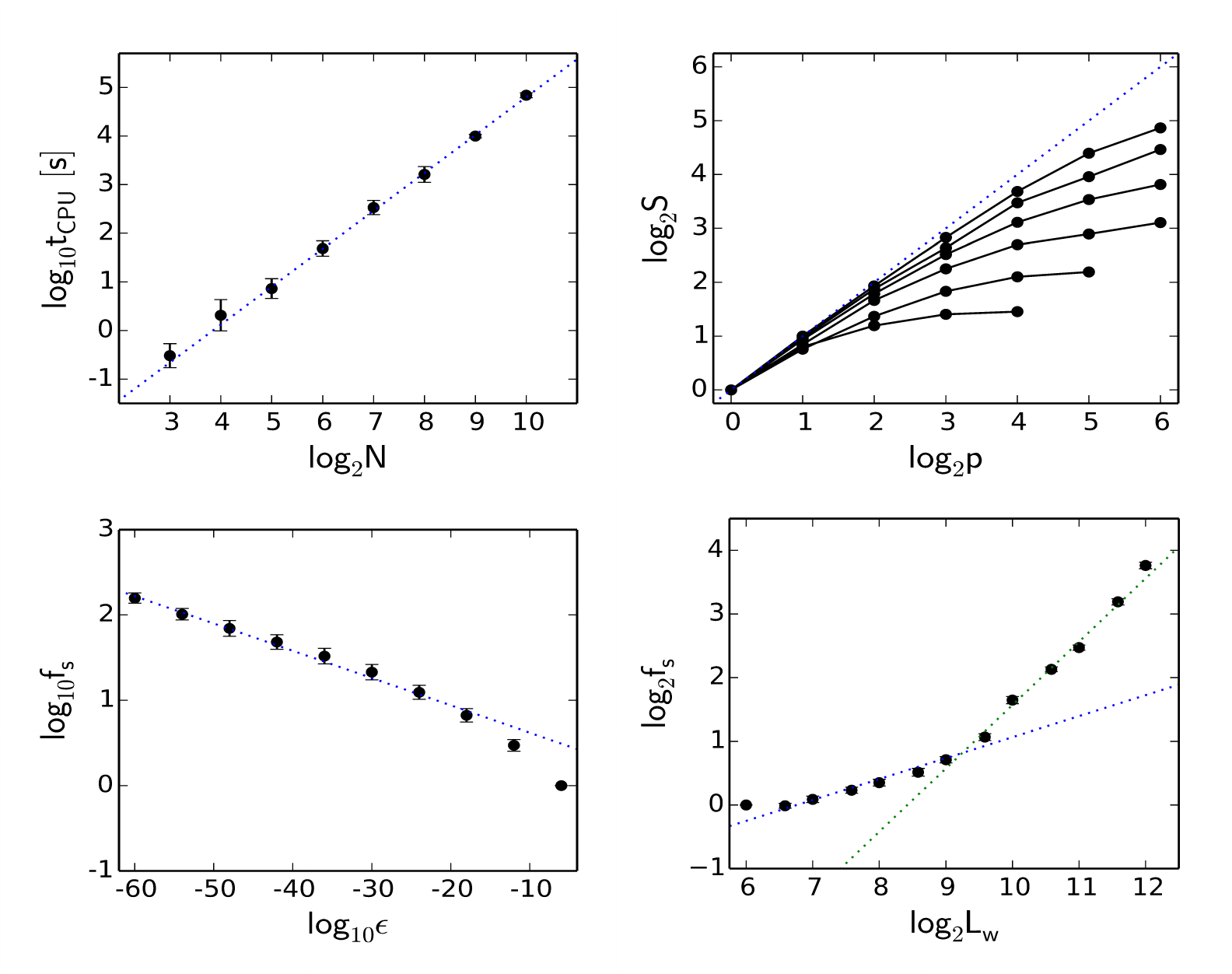} \\
\end{tabular}  
\caption{ \csentence{Scaling of \texttt{Brutus}.} 
In the top left panel we show the scaling of the wall-clock time that \texttt{Brutus} needs as a function of number of stars $N$. The dotted curve is a fit to the data given by t$_{\mathrm{CPU}} \propto N^{2.6}$.
In the top right panel we show the speed-up when the number of cores, p, is increased. The bottom, solid curve represents $N=32$ and each curve above has an $N$ a factor two higher than the previous curve. The dotted curve represents ideal scaling. In the bottom left panel we plot the slowdown factor as a function of the Bulirsch--Stoer tolerance $\epsilon$, for a fixed word-length of 1024\,bits. In the bottom right panel we plot the slowdown factor as a function of word-length $L_{w}$, for a fixed tolerance of 10$^{-10}$. The slowdown of the simulations is mainly caused by the very small Bulirsch-Stoer tolerances required.  }
\label{fig:scaling}
\end{figure*}

As a third test we simulate the dynamical formation of the first hard binary in a small star cluster. We select a moderate number of sixteen equal mass stars and draw them randomly from a Plummer distribution \citep{1911MNRAS..71..460P}. We integrate this system for about ten crossing times and apply the method of convergence. In Fig.~\ref{fig:plummer} we present how two solutions with the same initial conditions, but different precisions, diverge as a function of time. 
The rate of exponential divergence, on average, starts rather constant, with a loss of $\sim 2/3$ digits per time unit. This is equivalent to an e-folding time of $t_{e}=0.65$, which is consistent with the results of Goodman, Heggie and Hut (1993) (see their Fig.~8). From $t=20$ onwards, the rate of divergence experiences systematic changes, in particular a steep rise of the error of about 10 orders of magnitude between $t=26$ and $t=29$.  Such rises are a signature for the presence of a hard binary interacting with surrounding stars.  

The right panel in Fig.~\ref{fig:plummer} shows the energy conservation (black bullets, solid line) and the normalized phase space distance (red triangles, dashed line) versus $\epsilon$. Energy conservation is proportional to $\epsilon$, but the solutions only start to converge for $\epsilon<10^{-34}$. More generally, even if conserved quantities like total energy are conserved to machine-precision or better, it is not guaranteed that the solution itself has converged. 

The highest precision \texttt{Brutus} simulation in this example, ($\epsilon=10^{-50}$, $L_{w}=232$\,bits),  took about a day of wall-clock time, which is about 7000 times slower than a simulation with \texttt{Hermite} using $\eta=2^{-9}$. 

\subsection{Scaling of the wall-clock time}

The use of arbitrary-precision arithmetic dramatically increases the CPU time of \mbox{N-body} simulations. 
Also the BS method, which performs integration steps iteratively, makes an integration scheme more expensive by at least a factor two or more. To investigate for example how feasible it would be to run a converged \mbox{N-body} simulation for $10^3$ stars through core collapse, we perform a scaling test in which we vary the number of particles and the precision, $\epsilon$ and $L_{w}$. 

We randomly select positions and velocities for $N$ equal mass stars from the virialised Plummer distribution \citep{1911MNRAS..71..460P}, for $N=2, 4, 8$, ..., up to $1024$. The BS tolerance is fixed at a level of 10$^{-6}$ and the word-length at 64\,bits. We integrate the systems for one Hénon time unit and measure the wall-clock time. In the top left panel in Fig.~\ref{fig:scaling} we show the wall-clock time as a function of $N$, which fit the relation t$_{\mathrm{CPU}} \propto N^{2.6}$. 

For $N>32$, it becomes efficient to parallellise the code. Our version implements i-parallellisation \citep{2008Zwart} in the calculation of the accelerations. 
In the top right panel of Fig.~\ref{fig:scaling}, we plot the speed-up, $S$, against the number of cores. For  $N = 1024$, we obtain a speed up of a factor 30 using 64 cores.  

In the lower panels of Fig.~\ref{fig:scaling} we present the scaling of the wall-clock time with BS tolerance and word-length. To measure the dependence on BS tolerance, we simulated a 16-body cluster for 1 Hénon time unit. We varied the BS tolerance while keeping the word-length fixed at $L_{w} = 1024$\,bits. The relation obtained converges to t$_{\mathrm{CPU}} \propto \epsilon^{-0.032}$. A similar experiment was performed to measure the dependence on word-length. This time we fixed the BS tolerance at $\epsilon=10^{-10}$ and varied the word-length. For $L_{w} < 1024$, the relation can be estimated as $t_{CPU} \propto L_{w}^{0.33}$, while for $L_{w} > 1024$, $t_{CPU} \propto L_{w}$. This transition depends on the internal workings of the arbitrary-precision library which we will not discuss here. 

Using a very long word-length of 4096\,bits, i.e. $\sim10^{3}$ digits, results in a slowdown of a factor $f_{\mathrm{s}} \sim 16$ compared to 64\,bits. But for some simulations a BS tolerance smaller than 10$^{-50}$ can easily be required to reach convergence, and this will result in a slowdown of a factor $f_{\mathrm{s}} > 100$. The very small BS tolerance is often the main cause for the slowdown of the simulations, instead of the increased word-length. 

Using the above results, we can construct the following model to estimate the wall-clock time for integrating 1 Hénon time unit with $L_{w} < 1024$\,bits: 

\begin{equation}
t_{CPU} = \left( \frac{N}{512} \right)^{2.6} \left( \frac{\epsilon}{10^{-6} } \right)^{-0.032} \left( \frac{L_w}{64} \right)^{0.33} 10^{4}\,[s].  
  \label{eq:5}
\end{equation}

\noindent Integrating $N=1024$ with standard precision, ($\epsilon=10^{-6}$, $L_{w}=64$\,bits), up to core collapse at $\sim 300$ time units, and taking into account a speed up of a factor 30 due to parallellisation, we estimate a total wall-clock time of a week. Increasing the precision to ($\epsilon=10^{-20}$, $L_{w}=112$\,bits), will take about a month. A precision of ($\epsilon=10^{-50}$, $L_{w}=232$\,bits)
 will take roughly a year. 
To estimate how much precision is needed, we will assume that the rate of exponential divergence before the formation of the first hard binary is approximately constant. In the left panel of Fig.~\ref{fig:plummer}, the initial slopes correspond to a loss of $\sim 2/3$ digits per time unit. 
We construct the following approximate model for the initial BS tolerance needed to end up with a converged solution: 

\begin{equation}
\log_{10}{\epsilon} = \log_{10}{\delta_{\mathrm{final}}} - R_{\mathrm{div}} t_{\mathrm{cc}}.
  \label{eq:6}
\end{equation}

\noindent Here $\epsilon$ is the BS tolerance parameter,  $\delta_{final}$ is the final precision of all the coordinates in the system, $R_{\mathrm{div}}$ is the approximately constant rate of divergence, e.g. the number of accurate digits lost per unit of time, and $t_{\mathrm{cc}}$ is the core collapse time. We set the final precision to $10^{-6}$, i.e. convergence to the first 6 decimal places, and we set the core collapse time to $\sim 300$ as before. 
If we adopt $R_{\mathrm{div}} = 2/3$, we estimate that we need an $\epsilon \sim 10^{-206}$. This would take about $10^{5}$  years to finish. It would be more practical to simulate a 256-body cluster. If we set the core collapse to 100 time units we estimate $\epsilon \sim 10^{-73}$, which would take about a month on a cluster of 64 Intel(R) Xeon(R) E5530 cores. 

For direct \mbox{N-body} codes, the time for integrating up to core collapse usually scales as $\mathcal{O}(N^{3})$. Using the analysis above, we estimate that the time for converged core collapse simulations scales approximately exponentially. This is effectively caused by the exponential divergence.   

\section{Precision of statistical results: experimental setup}\label{Sect:Experiments}

\begin{figure*}[p]
\centering
\begin{tabular}{l}
\includegraphics[scale=0.98]{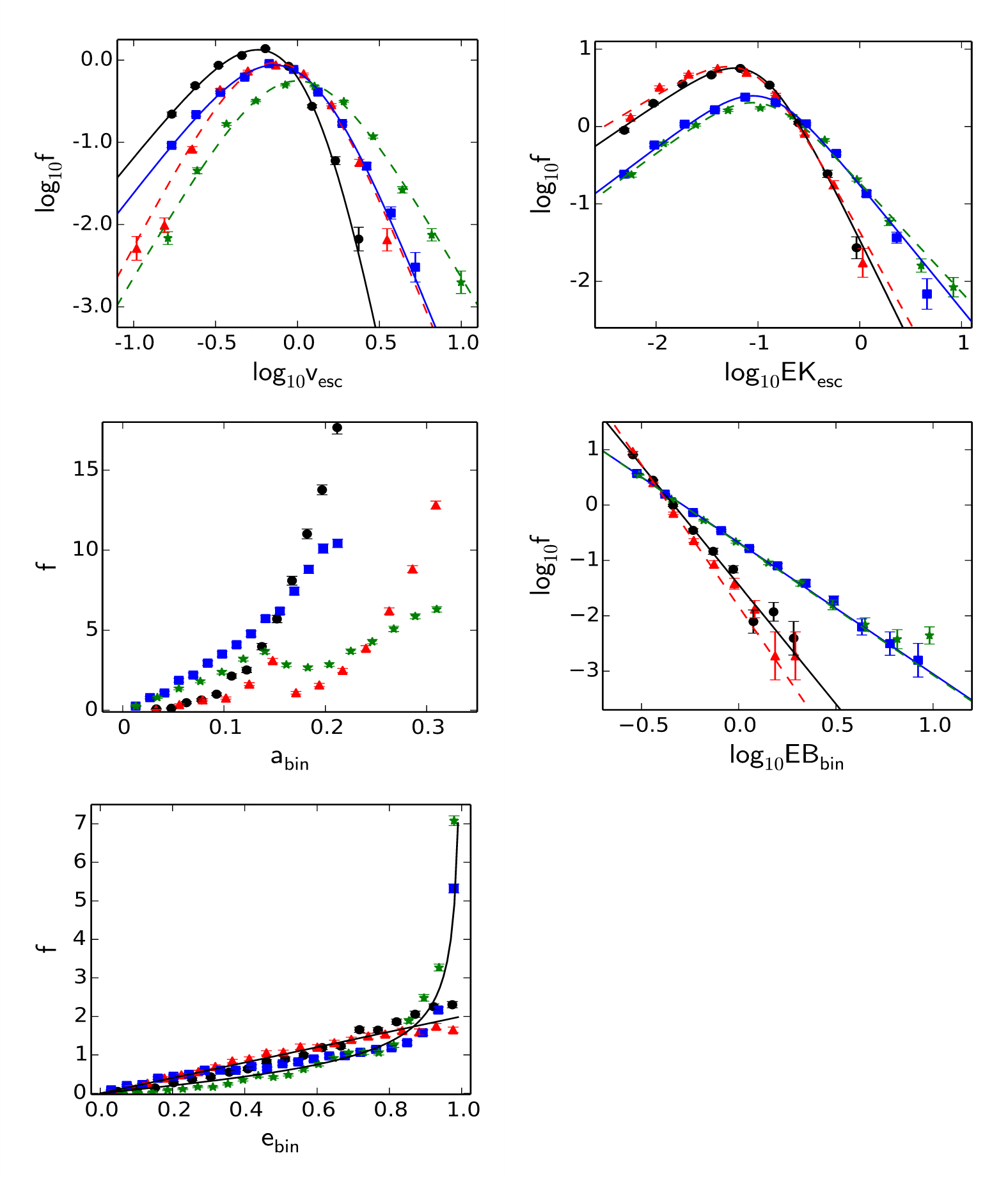} \\
\end{tabular}  
\caption{ \csentence{Comparison of \texttt{Brutus} results and analytical distributions. } 
Distributions are given for the escaper speed (top left) and kinetic energy (top right), binary semimajor axis (middle left),  binding energy (middle right) and binary eccentricity (bottom). The results from the \texttt{Brutus} simulations are represented by the data points, for each of the four sets of initial conditions: Plummer equal mass (black bullets), Plummer with different masses (red triangles), cold Plummer equal mass (blue squares) and cold Plummer with different masses (green stars).. Note that we use standard Hénon units \citep{1971Ap&SS..14..151H, 1986LNP...267..233H}.
Analytical models from the literature are fitted to the empirical distributions represented by the curves. For the eccentricities we plot the thermal distributions.  }
\label{fig:analytical}
\end{figure*}
In the previous section we demonstrated that it is possible to obtain a converged solution for a particular initial condition. 
We have also shown that a solution obtained by \texttt{Hermite} diverges from the converged solution, even up to the point that the microscopic solution given by \texttt{Hermite} is beyond recognition. 
We now perform a statistical study, to examine the hypothesis that double-precision \mbox{N-body} simulations 
produce statistically indistinguishable results, from those obtained from an ensemble of converged solutions with the same set of initial conditions. 
Because it is computationally expensive to reach convergence, we start 
investigating the hypothesis above by exploring the accuracy of 3-body statistics. 

The $N=3$ experiment is inspired by the Pythagorean problem, where after a complex 3-body interaction, a binary and an escaper are formed. As a variation to this, we define four different sets of initial conditions as follows: 

\begin{enumerate}
\item Plummer distribution equal mass
\item Plummer distribution with masses 1:2:4
\item Plummer distribution equal mass with zero velocities
\item Plummer distribution with masses 1:2:4 and zero velocities.
\end{enumerate}

\noindent The positions and velocities of the three stars are selected randomly from a virialised Plummer distribution \citep{1911MNRAS..71..460P, 1974A&A....37..183A}. For the cold collapse systems, we set the velocities to zero. Then we rescale the positions and velocities to virialise the systems if the initial velocities are non-zero, or we set the total energy equal to $E=-0.25$ if the system starts out cold. We adopt standard Hénon units \citep{1971Ap&SS..14..151H, 1986LNP...267..233H} throughout.

In the case of the cold initial conditions, the systems start democratically, i.e. the minimal distance between each pair of particles is greater than $N^{-1}$. We reject initial conditions in which this criterion is not satisfied. 
This is to prevent initial realisations where two stars which are very near, fall to each other radially causing very long wall-clock times for the integration. 
When starting with a democratic configuration, there will also be an initial close triple encounter \citep{1994CeMDA..60..131A}, which is hard to integrate accurately
and is therefore a good test. 
A total of 10000 random realisations are generated for each set of initial conditions and can be found in the accompanying data files. 

We stop the simulations when the system is dissolved into a permanent binary and an escaper.
The criteria used to detect an escaper are the following: 

\begin{enumerate}
\item escaper has a positive energy, $E >0$,
\item is a certain distance away from the center of mass, $r > 2\,r_{\mathrm{virial}}$,
\item is moving away from the center of mass, $r \cdot v > 0$,
\end{enumerate}

\noindent The energy of the escaper is calculated in the barycentric frame of the three particles and $r_\mathrm{virial}$ is the virial radius of the system, which is of the order unity in Hénon units. 

There may be situations in which a star is ejected without actually escaping from the binary. After a long excursion the star turns around and once again engages the binary in a 3-body resonance \citep{1983ApJ...268..319H}. 
Because these systems need to be integrated for a longer time, they also require higher precision to reach convergence, which takes a long time to integrate \citep[see also][]{1993ApJ...403..256H}. 
To deal with this issue, we perform the simulations iteratively by increasing the final integration time t$_{\mathrm{end}}$. Starting with $t_{\mathrm{end}}=50$ Hénon time units, we evolve every system and detect those that are dissolved. Then we increase t$_{\mathrm{end}}$ to 100, 150, 200 etc., but only for those systems which have not yet dissolved.  
A complete ensemble of solutions is obtained up to t$_{end}$ $\sim$ 500, or equivalently $\sim$ 180 crossing times where the crossing time has a value of $2\sqrt{2}$ in Hénon units \citep{1971Ap&SS..14..151H, 1986LNP...267..233H}. Systems which take a longer time to integrate are not taken into account in this research. The fraction of long-lived systems is however a statistic we measure.  We gathered the final, converged configurations in the accompanying data files.

Each initial realisation is run with the \texttt{Hermite} code, using standard double-precision, 
and with \texttt{Brutus}, using arbitrary-precision until a converged solution is obtained. 
At the end of each simulation, we investigate the nature of the binary and the escaper. 
In addition to the BS tolerance, word-length, CPU time and dissolution time, we record the mass, speed and escape direction of the escaping single star, and the semimajor axis, binding energy and eccentricity of the binary. In this way, we obtain statistics for $N=3$ generated by a conventional \mbox{N-body} solver and by \texttt{Brutus}.

\section{Results}\label{Sect:Results}

Before we perform a detailed comparison between results obtained by \texttt{Hermite} and \texttt{Brutus}, we first compare the \texttt{Brutus} results with analytical distributions from the literature in order to relate to previous studies. 
We compare \texttt{Hermite} and \texttt{Brutus} on a global level by performing two-sample Kolmogorov--Smirnov tests \citep{1933Kolmogorov, 1948Smirnov} to see whether global distributions are statistically indistinguishable. 
We also compare the distribution of lifetimes of triples to see whether precision influences the stability and we measure the typical CPU time and BS tolerance needed to obtain a converged solution. 
After this, we compare \texttt{Hermite} and \texttt{Brutus} per individual system, with the aim of investigating the nature of the differences of every individual outcome. 
Finally, we define categories which classify a conventional simulation as a preservation or exchange, depending on whether the identity of the escaping star is consistent between \texttt{Hermite} and \texttt{Brutus}.  

\subsection{\texttt{Brutus} versus analytical distributions}

\begin{table}[t]
\begin{tabular}{p{3.5cm} p{0.25cm} p{0.05cm} p{0.4cm} p{0.25cm} p{0.05cm} p{0.4cm}}
\hline
Velocity & $\alpha$ & & & $\beta$ & & \\
\hline
Plummer equal mass         &  2.5  & $\pm$ & 0.09 & 6.7 & $\pm$ & 1.02 \\
Plummer mass ratio           & 3.8  & & 0.16 & 4.4 & & 0.43 \\
Cold Plummer equal mass &  2.6 & & 0.19 & 3.8 & & 0.28 \\
Cold Plummer mass ratio  &  3.4  & & 0.45 & 3.4 & & 0.19 \\
\hline
Kinetic energy & & & & & & \\
\hline
Plummer equal mass         &  0.9 & & 0.02 & 1.8 & & 0.04 \\
Plummer mass ratio           & 0.8  & & 0.02 & 1.6 & & 0.04 \\
Cold Plummer equal mass &  0.99 & & 0.02 & 1.3 & & 0.03 \\
Cold Plummer mass ratio  &  0.98  & & 0.03 & 1.2 & & 0.02 \\
\hline
Binding energy & & & & & & \\
\hline
Plummer equal mass          & 4.31  & & 0.13  &  & & \\
Plummer mass ratio           & 5.12  & & 0.32 & & & \\
Cold Plummer equal mass & 2.37   & & 0.11  &  & & \\
Cold Plummer mass ratio   & 2.38  & & 0.12  & & & \\
\hline
\end{tabular}
\caption{Fitted powerlaw indices for the velocity and kinetic energy distributions of the escaping stars and for the binding energy distribution of the binary stars. Note that we use equal intervals in logarithmic space. }
\label{tablev}

\end{table}

\begin{figure*}[t]
\centering
\begin{tabular}{l}
\includegraphics[scale=0.88]{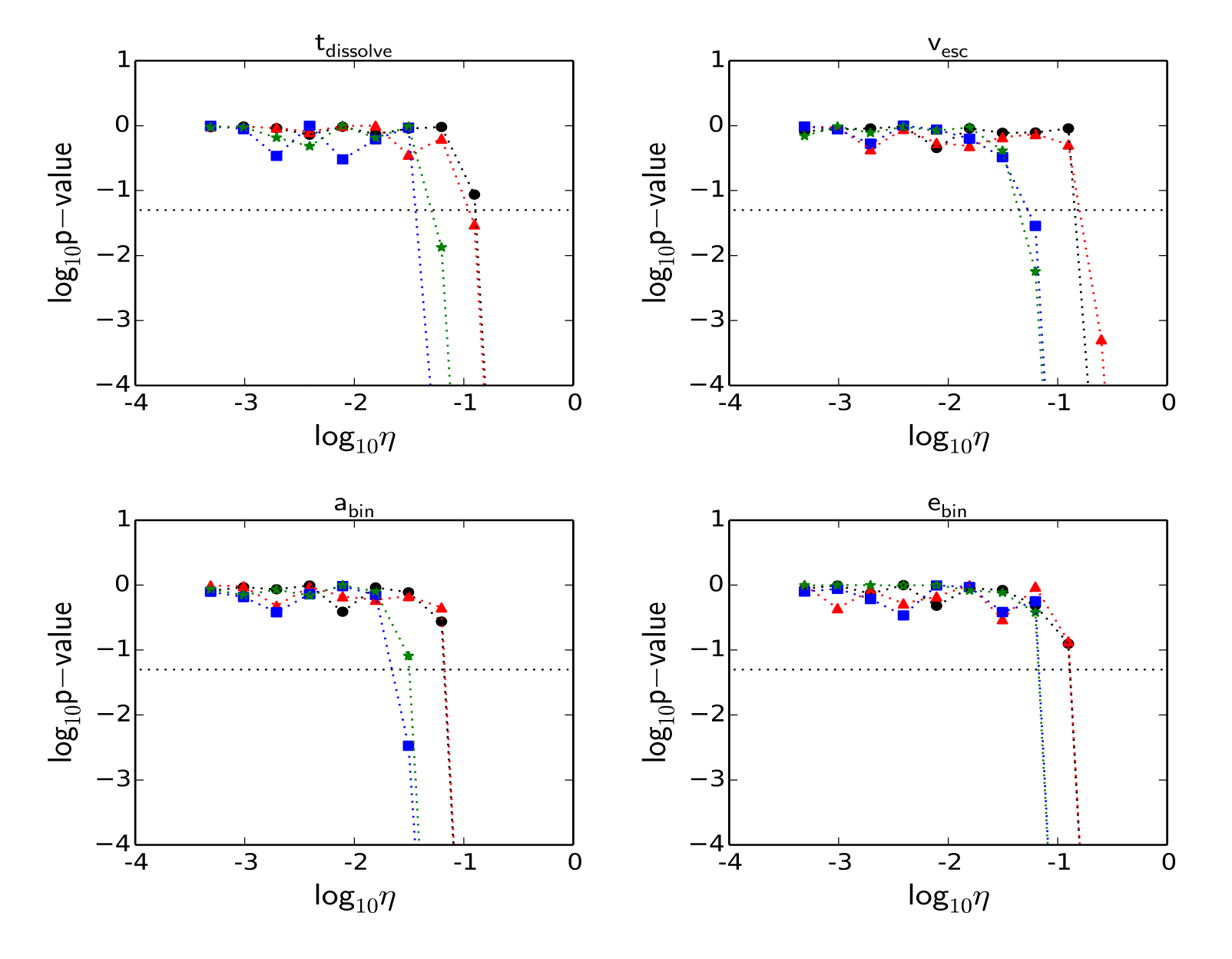} \\
\end{tabular}  
\caption{ \csentence{Two-sample K--S tests on distributions obtained by \texttt{Hermite} and \texttt{Brutus}.} 
We compare distributions of dissolution time (top left), escaper speed (top right), binary semimajor axis (bottom left) and binary eccentricity (bottom right)). The color coding is the same as in Fig.~\ref{fig:analytical}. Two-sample K--S tests are performed and the p--value is plotted versus \texttt{Hermite} time-step parameter $\eta$. 
The dashed line represents the 5$\%$ significance level. For $\eta < 2^{-5}$, the distributions are not significantly different. }
\label{fig:ks}
\end{figure*}

In Fig.~\ref{fig:analytical}, the distributions obtained by converged solutions are given for the following quantities: velocity and kinetic energy of the escaper in the barycentric reference frame, and semimajor axis, binding energy and eccentricity of the binary. 
We start by looking at the eccentricity distributions (bottom panel in Fig.~ \ref{fig:analytical}). These distributions can be estimated analytically by assuming that the probability of a certain configuration is proportional to the associated volume in phase space \citep{1976MNRAS.176...63M, 2006Valtonen} or by considering an equilibrium distribution of binary stars in a cluster \citep{1975MNRAS.173..729H}. The resulting thermal distribution in the three-dimensional case is given by  

\begin{equation}
f(e) = 2e,  
  \label{eq:7}
\end{equation}

\noindent and in the two-dimensional case by 

\begin{equation}
f(e) = \frac{e}{\sqrt{1-e^{2}}}, 
  \label{eq:8}
\end{equation}

\noindent The 3-body cold collapse problem is essentially a two-dimensional problem. We compare the empirical and theoretical distributions by means of the K--S test (see also next section). It turns out that the distributions in eccentricity are statistically distinguishable. By inspection by eye we observe that in the virialised case, there are slight deviations at high eccentricities. In the case of the equal-mass, cold systems, there are more low eccentricity binaries compared to the theoretical prediction. They coincide at an eccentricity of about 0.7, after which they deviate again. For the cold systems with unequal masses, this behaviour is the other way around. The analytical predictions are able to capture the empirical distributions only in a qualitative manner.  

The velocity distribution of the single escaping star can be estimated analytically in a similar way as was done for the eccentricities. The resulting distribution is predicted to be a double powerlaw given by \citep{1976MNRAS.176...63M, 2006Valtonen}:

\begin{equation}
f(v) \propto \frac{v^{\alpha}}{(1 + \gamma v^{2})^\beta}.   
  \label{eq:9}
\end{equation}

\noindent We fit this model to the data (see Fig.~\ref{fig:analytical}, first panel) and obtain values for $\alpha$ and $\beta$ which are given in Table~\ref{tablev}. The powerlaw indices vary with mass ratio and total angular momentum. 
To remove the dependence on mass ratio, we plot the kinetic energy of the escaper (see Fig.~\ref{fig:analytical}, top right panel). Again, we fit a double powerlaw of a similar form as Eq.~\ref{eq:9}, and the powerlaw indices are given in Table~\ref{tablev}. Both the escaper velocity and kinetic energy are consistent with a double powerlaw distribution. 

The binary semimajor axis and binding energy are related quantities. We fit the binding energy distribution (see Fig.~\ref{fig:analytical}, middle right panel) to a powerlaw \citep{1975MNRAS.173..729H, 1976MNRAS.176...63M, 2006Valtonen}: 

\begin{equation}
f(E_{\mathrm{B}}) \propto E_{\mathrm{B}}^{-\alpha}. 
  \label{eq:10}
\end{equation}

\noindent The fitted powerlaw indices are given in Table~\ref{tablev}. The empirical distributions are consistent with a powerlaw, although somewhat steeper than predicted \citep{1975MNRAS.173..729H, 1976MNRAS.176...63M, 2006Valtonen}. The slopes do tend to vary somewhat as a function of angular momentum \citep{1976MNRAS.176...63M, 2006Valtonen}. 




The empirical distributions obtained by \texttt{Brutus} are in qualitative agreement with the analytical estimates present in the literature \citep{1975MNRAS.173..729H, 1976MNRAS.176...63M, 2006Valtonen}. Slight variations are present due to the dependence on total angular momentum, a limited statistical sampling and assumptions made in the derivation of the analytical distributions. Nevertheless, a similar qualitative agreement has been obtained between the analytical distributions discussed above and empirical distributions from an ensemble of conventional numerical solutions, e.g. not converged \citep[][chapters 7--8 and references therein]{2006Valtonen}. The question remains to what extend conventional and converged solutions agree quantitatively. 

\subsection{\texttt{Brutus} versus \texttt{Hermite}: global comparison}\label{K_Stest}

A quantitative way to compare global distributions is by performing two-sample Kolmogorov--Smirnov tests (K--S tests) \citep{1933Kolmogorov, 1948Smirnov}. 
The K--S test gives the likelihood that two samples are drawn from the same distribution, quantified by the value called $p$. 
When the $p$-value is below five percent, the distributions are considered to be significantly different. 

In Fig.~\ref{fig:ks} we plot the $p$-value obtained by comparing the \texttt{Brutus} distribution with the \texttt{Hermite} distribution versus time-step parameter $\eta$ used for \texttt{Hermite}.
In the panel showing the data for the binary semimajor axis, the distributions of the cold systems become significantly different for $\eta > 2^{-6}$. The distributions from the initially virialised systems start to differ for $\eta > 2^{-4}$. 
The cold systems are harder to model accurately, because of the close encounters that occur shortly after the start. 
The reason the distributions start to become significantly different at large time-steps is because at these large time-steps most simulations violate energy conservation by $\left| \Delta E/E \right| > 0.1$. When this occurs, solutions might reach regions in $6N$-dimensional phase-space, which theoretically are forbidden. The distribution then becomes biased by these outlier solutions. 


\subsection{Lifetime of triple systems}

\begin{figure}[t]
\centering
\begin{tabular}{l}
\includegraphics[scale=0.77]{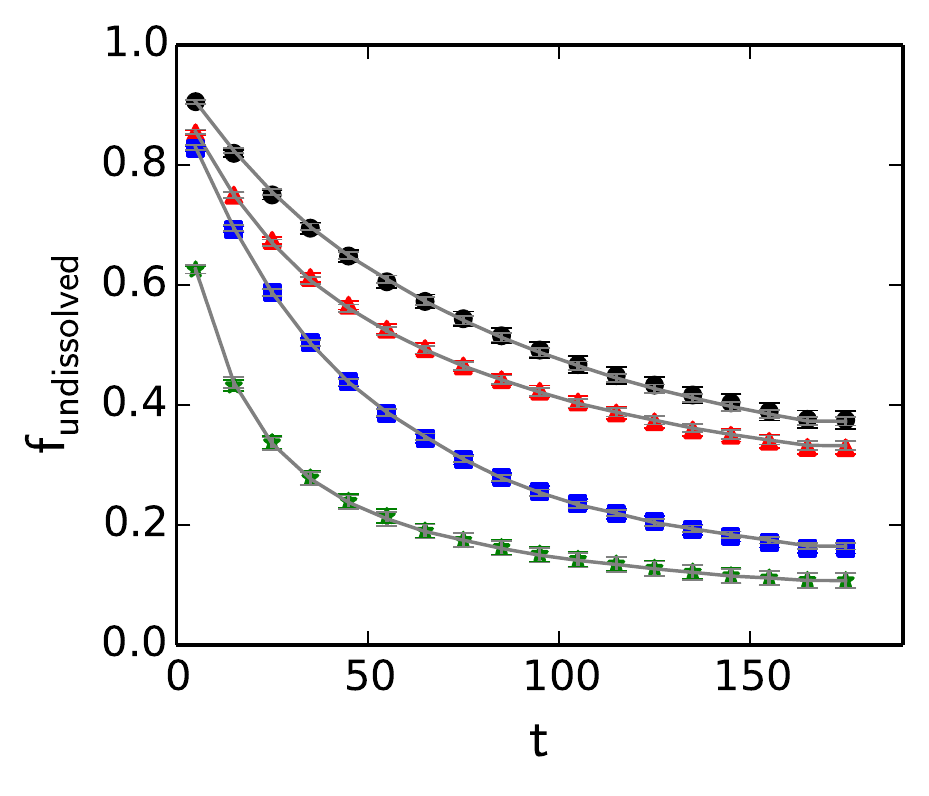} \\
\end{tabular}  
\caption{  \csentence{Lifetime of triple systems.} 
We plot the fraction of triple systems that have not dissolved yet into a permanent binary and escaping single star configuration, as a function of simulation time (in units of crossing time). The color coding is the same as in Fig.~\ref{fig:analytical}. The grey curves through the data points represent the interpolated \texttt{Hermite} results with a time-step parameter $\eta = 2^{-5}$.  }
\label{fig:fundissolved}
\end{figure}

\begin{figure*}
\centering
\begin{tabular}{l}
\includegraphics[scale=0.88]{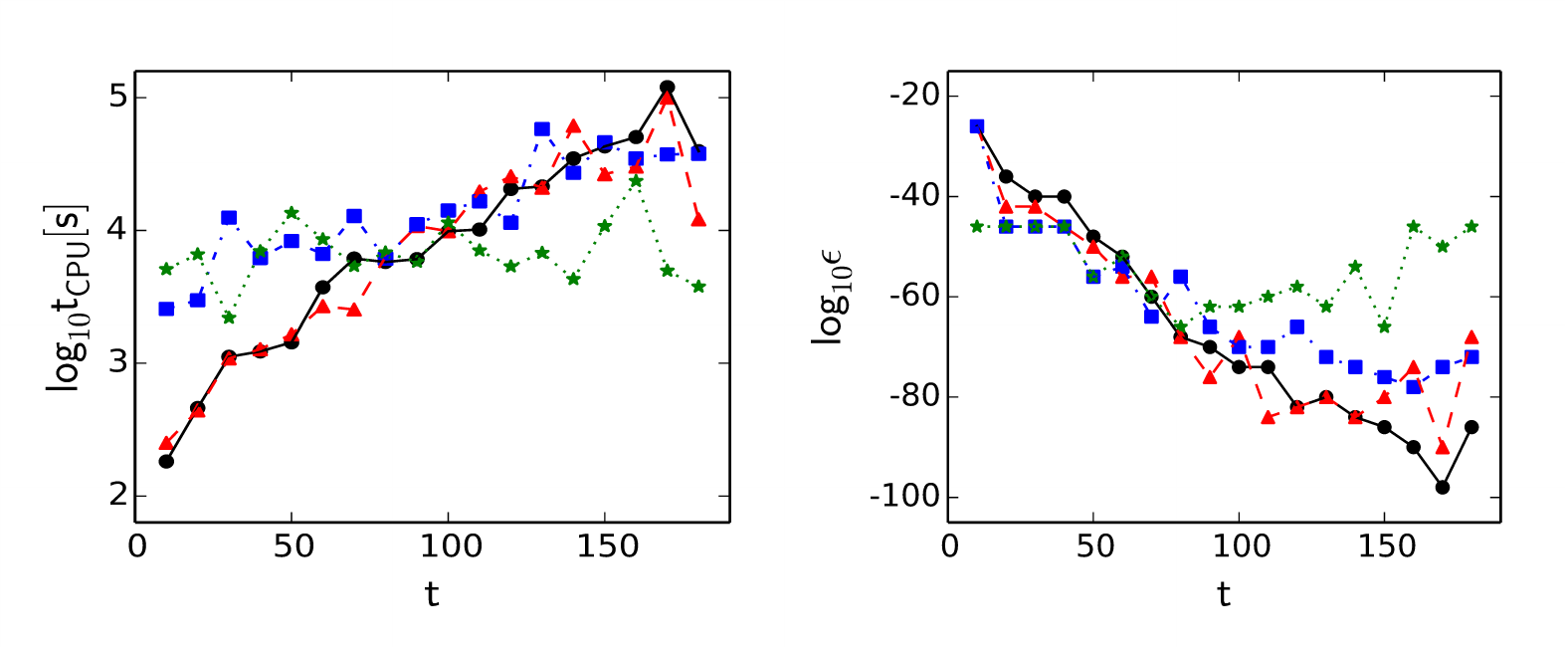} \\
\end{tabular}  
\caption{  \csentence{CPU time and precision as a function of  time for \texttt{Brutus}. } 
On the left, we plot the CPU time of the simulation which took the longest, as a function of dissolution time. On the right, we plot the Bulirsch--Stoer tolerance of the simulation which needed the highest precision, as a function of dissolution time. The different curves represent the four sets of initial conditions as in the previous plots.  }
\label{fig:ensemble}
\end{figure*}

In Fig.~\ref{fig:fundissolved}, we present the fraction of triple systems which are undissolved, i.e. still interacting, as a function of time. The results by \texttt{Brutus} are represented by the data points: equal-mass Plummer (black bullets), Plummer with different masses (red triangles), equal-mass cold Plummer (blue squares) and cold Plummer with different masses (green stars). The results by \texttt{Hermite} for a time-step parameter $\eta = 2^{-5}$ are represented by the curves appearing to go through the data points. 

The initially cold systems dissolve faster than the initially virialised systems. This is somewhat expected due to the close triple encounter resulting from the initial cold collapse: the rate of energy exchange can be very high for these encounters \citep{1991PASP..103..359J}. After $\sim$ 180 crossing times, about 40$\%$ of the systems which started with an equal-mass Plummer initial configuration, are undissolved, compared to about 10$\%$ for the cold Plummer with different masses.  Systems which include stars with different masses dissolve faster than their equal mass counterparts. Energy equipartition tends to cause the lightest particle to quickly reach the escape velocity. 

In Fig.~\ref{fig:fundissolved}, the grey curves through the data points represent the interpolated \texttt{Hermite} results.  Even though \texttt{Hermite} and \texttt{Brutus} use different algorithms and precisions to solve the equations of motion, we find that  the lifetime of an unstable triple is statistically indistinguishable between converged \texttt{Brutus} and non-converged \texttt{Hermite} solutions (but see also Sec.~\ref{MacroProp}). 

  

In Fig.~\ref{fig:ensemble}, we plot the maximum CPU time and minimum BS tolerance, both as a function of dissolution time. This is shown for the \texttt{Brutus} simulations, for the four different initial conditions. The longer it takes for a system to dissolve, the longer the CPU time and the higher the precision needed to reach a converged solution. To reach $\sim$180 crossing times, there are systems which require a BS tolerance of the order $10^{-100}$, with the final converged run taking of the order a few days. The average CPU time as a function of time is about an order of magnitude smaller than the maximum CPU time. The average BS tolerance ranges from $\sim 10^{-20}$ to $10^{-30}$. 
For systems which dissolve within 100 crossing times, \texttt{Brutus} is on average about a factor 120 slower than \texttt{Hermite}. 

We were able to obtain a complete ensemble of systems dissolving within $\sim 180$ crossing times. Simulations which take longer than this are not taken into account in this experiment. The fraction of long-lived systems as obtained by \texttt{Hermite} and \texttt{Brutus} are consistent. For our purpose of comparing results from conventional integrators with the converged solution, integrating up to $\sim 180$ crossing times is sufficient, in the sense that there is enough time for conventional solutions to diverge from the true solution (see Sec.~\ref{sec:facc}). Including the long-lived triple systems may however influence the statistical distributions and biases on the long term. 

\subsection{\texttt{Brutus} versus \texttt{Hermite}: individual comparison}

\begin{figure*}[t]
\centering
\begin{tabular}{l}

\includegraphics[scale=0.95]{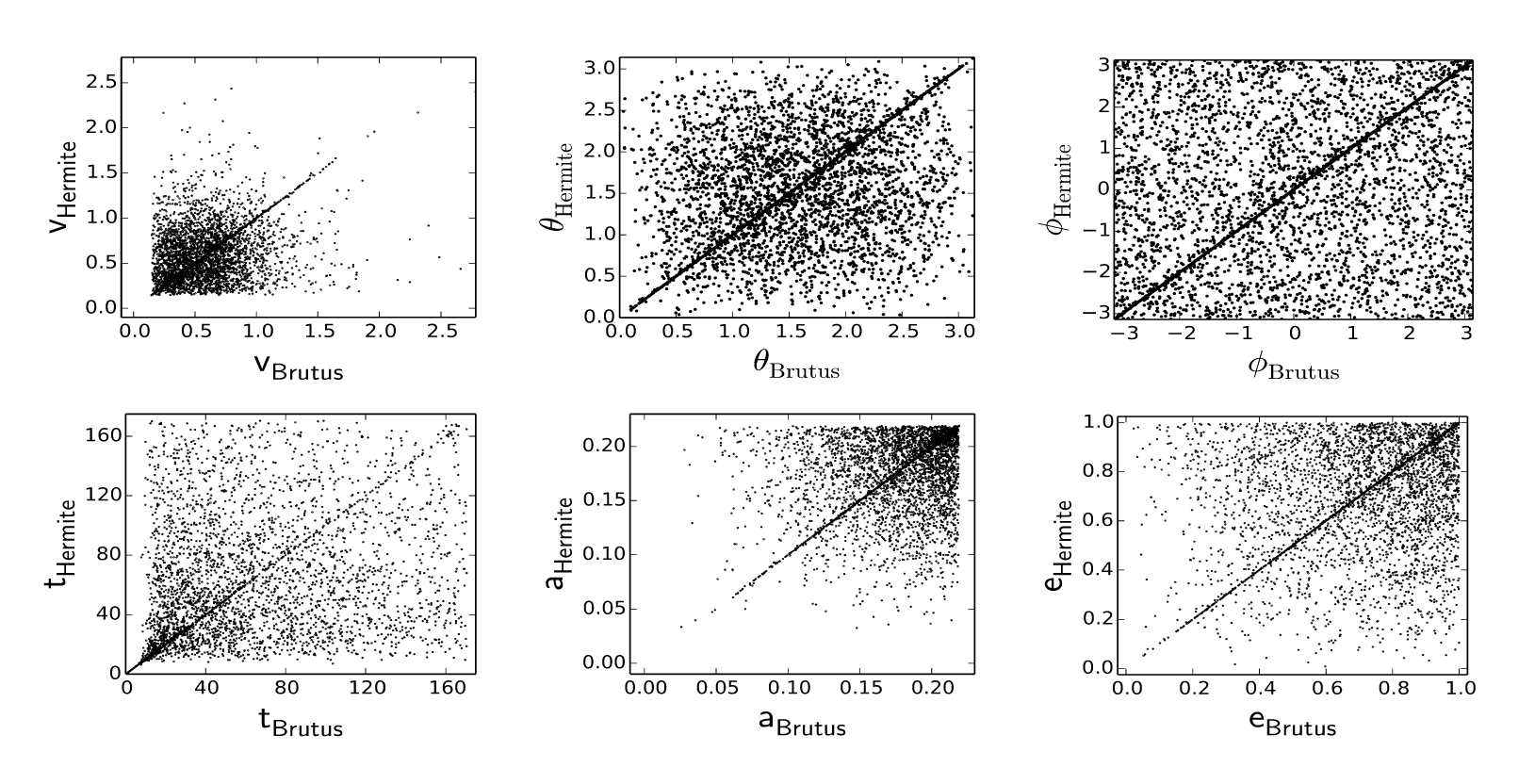} \\

\end{tabular}  
\caption{ \csentence{Direct comparison of \texttt{Brutus} and \texttt{Hermite} results per individual simulation. } 
The results are shown only for the $N=3$ equal mass Plummer data set and for a \texttt{Hermite} time-step parameter $\eta=2^{-5}$. Each dot in a panel represents a different initial realisation. The value on the ordinate is the value obtained using \texttt{Hermite} and the value on the abscissa the value obtained by \texttt{Brutus}. We compare the escaper velocity (top left), direction of the escaper: polar angle (top middle) and azimuthal angle (top right), (with respect to the plane of the binary and pericentre direction), dissolution time (bottom left), binary semimajor axis (bottom middle) and binary eccentricity (bottom right). The diagonal represents accurate \texttt{Hermite} solutions. The scatter around it represents solutions where \texttt{Hermite} and \texttt{Brutus} have diverged. }
\label{fig:scatter}
\end{figure*}

For the individual comparison, we take a certain initial realisation and compare the solutions of \texttt{Hermite} and \texttt{Brutus}. In Fig.~\ref{fig:scatter} we show scatter plots of the \texttt{Hermite} solution (with time-step parameter $\eta=2^{-5}$) versus the converged \texttt{Brutus} solution for the equal-mass Plummer data set. 

Data points on the diagonal represent accurate solutions, whereas the scatter around it represents inaccurate \texttt{Hermite} solutions. The diagonal is present in each panel and extends throughout the range of possible outcomes. The width of the diagonal is very narrow. When the normalized phase-space distance between the \texttt{Hermite} and \texttt{Brutus} solution $\delta < 10^{-1}$, then the coordinates are accurate enough to produce derived quantities accurate to at least one decimal place and \texttt{Hermite} and \texttt{Brutus} will give similar results. Once $\delta > 10^{-1}$, the solution has diverged to a different trajectory in phase-space leading to a different outcome. This outcome could in principle be any of the possible outcomes as can be derived from the amount of scatter in the \texttt{Hermite} solutions at a fixed \texttt{Brutus} solution.   

In the scatter plot of the dissolution time, we observe that for small times ($t < 10$), \texttt{Hermite} and \texttt{Brutus} agree on the solution in the sense that the data points lie on the diagonal. Systems which dissolve after a short time don't have sufficient time to accumulate enough error to diverge to another trajectory in phase-space. Once however this level of divergence is reached, the scatter immediately covers the entire, available outcome space. This randomisation is also observed in the other panels. 

\subsubsection{The fraction of accurate solutions}\label{sec:facc}

\begin{figure}
\centering
\begin{tabular}{l}

\includegraphics[scale=0.78]{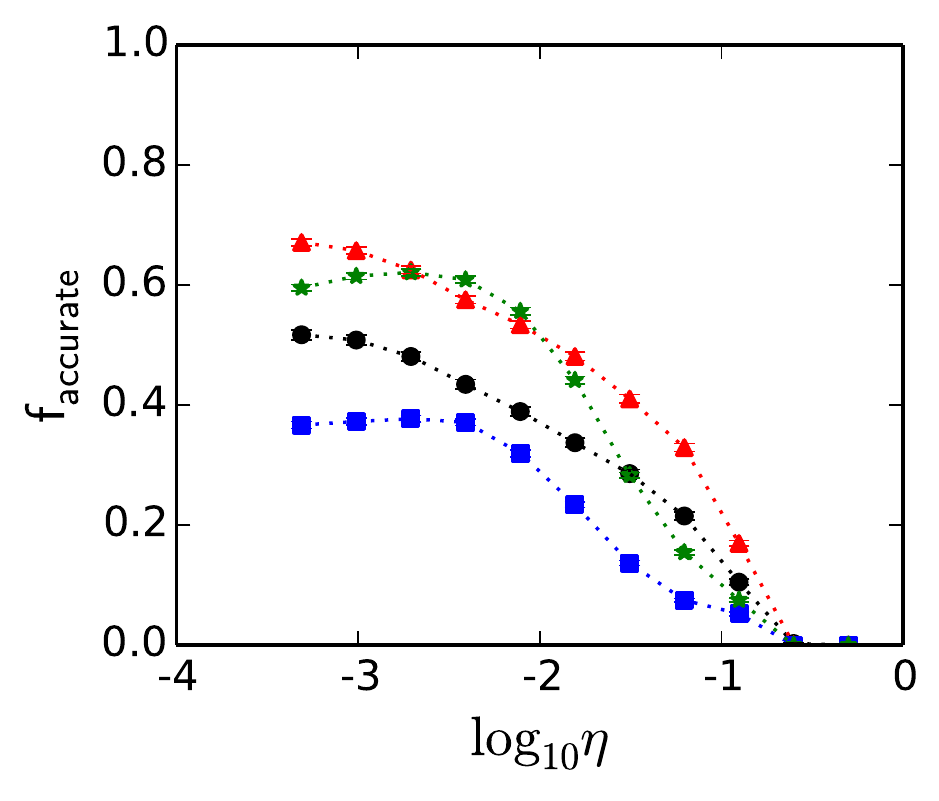} \\

\end{tabular}  
\caption{ \csentence{The fraction of accurate \texttt{Hermite} simulations as a function of \texttt{Hermite} time-step parameter $\eta$.} 
  The different curves represent the different data sets: equal mass Plummer (black bullets), Plummer with different masses (red triangles), equal mass cold Plummer (blue squares) and cold Plummer with different masses (green stars). As $\eta$ decreases, the accurate fraction increases. However, for $\eta < 2^{-7}$, the fraction starts to saturate, more so for the cold data sets. At this point the effect of round-off error becomes important.   }
\label{fig:facc}
\end{figure}

In Fig.~\ref{fig:facc} we estimate the fraction of data points on the diagonal as a function of the \texttt{Hermite} time-step parameter, $\eta$. 
We only include the data points for which the normalized phase-space distance $\delta < 10^{-1}$. 
For the largest time-step parameters used ($\eta > 10^{-1}$) the fraction on the diagonal, or the accurate fraction, varies from zero to about 0.2. By reducing the time-step parameter, the accurate fraction increases until it saturates at about 0.4 to 0.7 depending on the initial conditions. Even though by reducing $\eta$, the discretisation error decreases, the number of integration steps increases, which then increases the round-off error. 
For the data sets with zero angular momentum, the maximum accurate fraction is obtained for $\eta \sim 2^{-9}$. For the initially virialised systems this seems to occur between $\eta \sim 10^{-3}-10^{-4}$, although the actual saturation point is not visible yet. This dependence on angular momentum is due to the initial cold collapse and subsequent close encounters, which increases the round-off error.  

\subsubsection{The error distribution}

\begin{figure*}[t]
\centering
\begin{tabular}{l}

\includegraphics[scale=0.95]{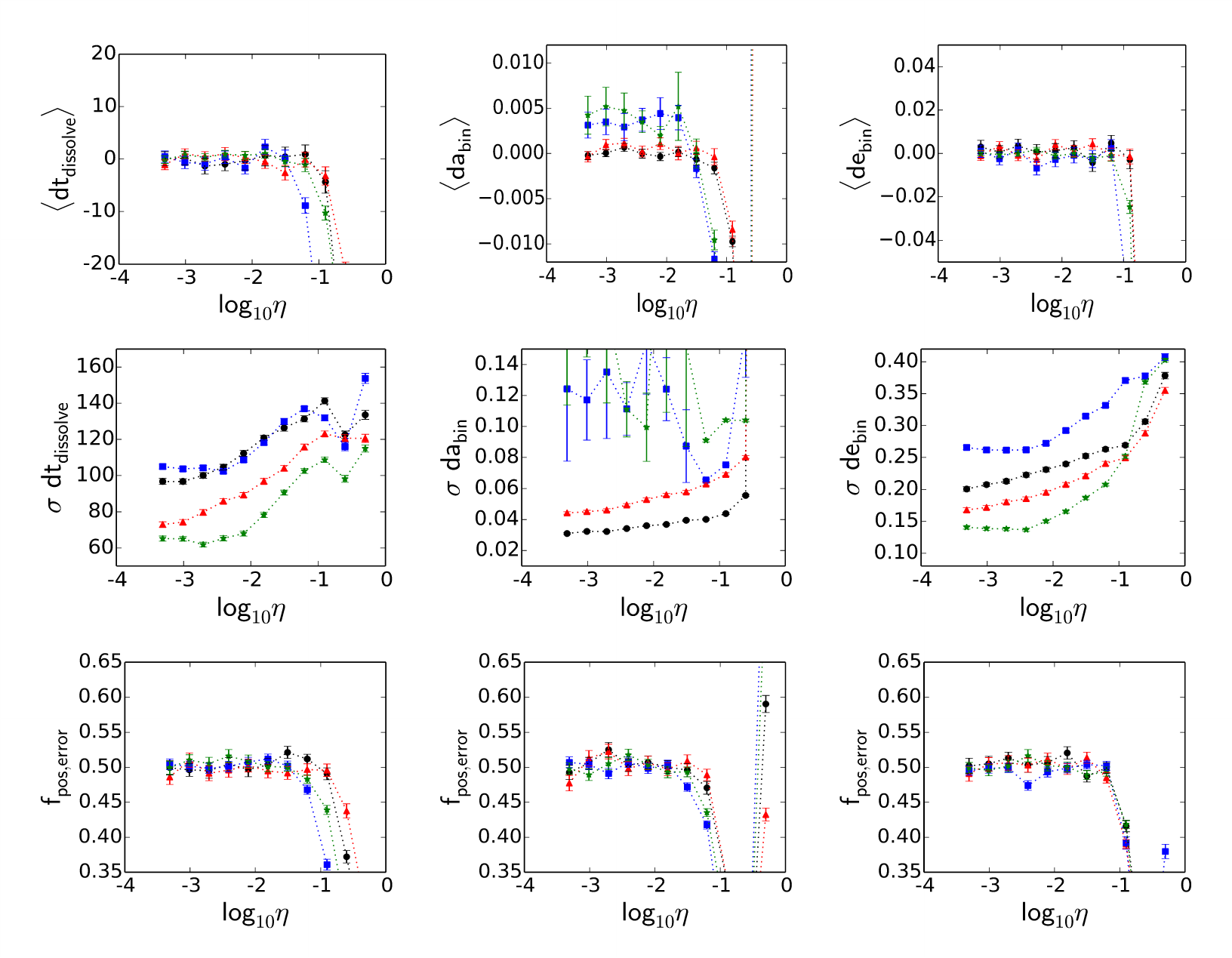} \\

\end{tabular}  
\caption{  \csentence{Statistics on the error distribution of \texttt{Hermite} results.} 
We present the average error (top row), the standard deviation of the error distribution (middle row) and the fraction of errors which are positive (bottom row). The errors are given for the dissolution time (left column), binary semimajor axis (middle column) and eccentricity (right column). The different curves represent the different data sets similar as in Fig.~\ref{fig:facc}. }
\label{fig:error}
\end{figure*}

In Fig.~\ref{fig:error} we present statistics on the distribution of the errors, i.e. $S_{\texttt{Hermite}}-S_{\texttt{Brutus}}$, with S a statistic. 
For the dissolution time and the eccentricity, the average error converges to zero for $\eta < 10^{-1}$. For larger time-steps, simulations which grossly violate energy conservation ($\left| \Delta E/E \right| > 0.1$) cause biases in the average error. For the binary semimajor axis however, the data representing the cold collapse simulations also seem to be systematically biased for small time-steps, in the sense that \texttt{Hermite} makes fewer tight binaries. 

The width of the error distributions converge to a non-zero value. This can be understood because with decreasing time-step, round-off errors will become more important so that the standard deviation of the errors will never reach zero. For the dissolution time, the width of the error distribution for the smallest time-step parameter adopted, varies from 60 to 100 crossing times. For the eccentricities the width is on average $\sim 0.2$. For the semimajor axis the width approaches $\sim 0.05$ (in Hénon units). In the case of the semimajor axis, the data representing the cold collapse simulations behave differently, because the width is much larger than the width for the data representing the initially virialised systems. 

If we regard the results given by \texttt{Brutus} and \texttt{Hermite} as random variables drawn from the same distribution, then we can write the variance in a certain statistic, in this example the eccentricity, as:

\begin{equation}
\langle (e_H-e_B)^2 \rangle = \langle e_H^2 \rangle + \langle e_B^2 \rangle - 2\langle e_H \rangle \langle e_B \rangle.
  \label{eq:15}
\end{equation}

\noindent Here $e$ stands for eccentricity and the subscripts for \texttt{Brutus} and \texttt{Hermite}. For a thermal eccentricity distribution (Eq. \ref{eq:7}), we obtain a standard deviation of $1/3$. However, this only applies to inaccurate \texttt{Hermite} results, which had enough time to diverge through outcome space. If we multiply the theoretical standard deviation calculated above by the inaccurate fraction, we obtain a range in the standard deviation from 0.17 to 0.27, as $\eta$ ranges from the most precise value to $\eta = 10^{-1}$. 

\subsubsection{Symmetry of the error distribution}\label{symresult}

\begin{figure*}[t]
\centering
\begin{tabular}{l}
\includegraphics[scale=0.88]{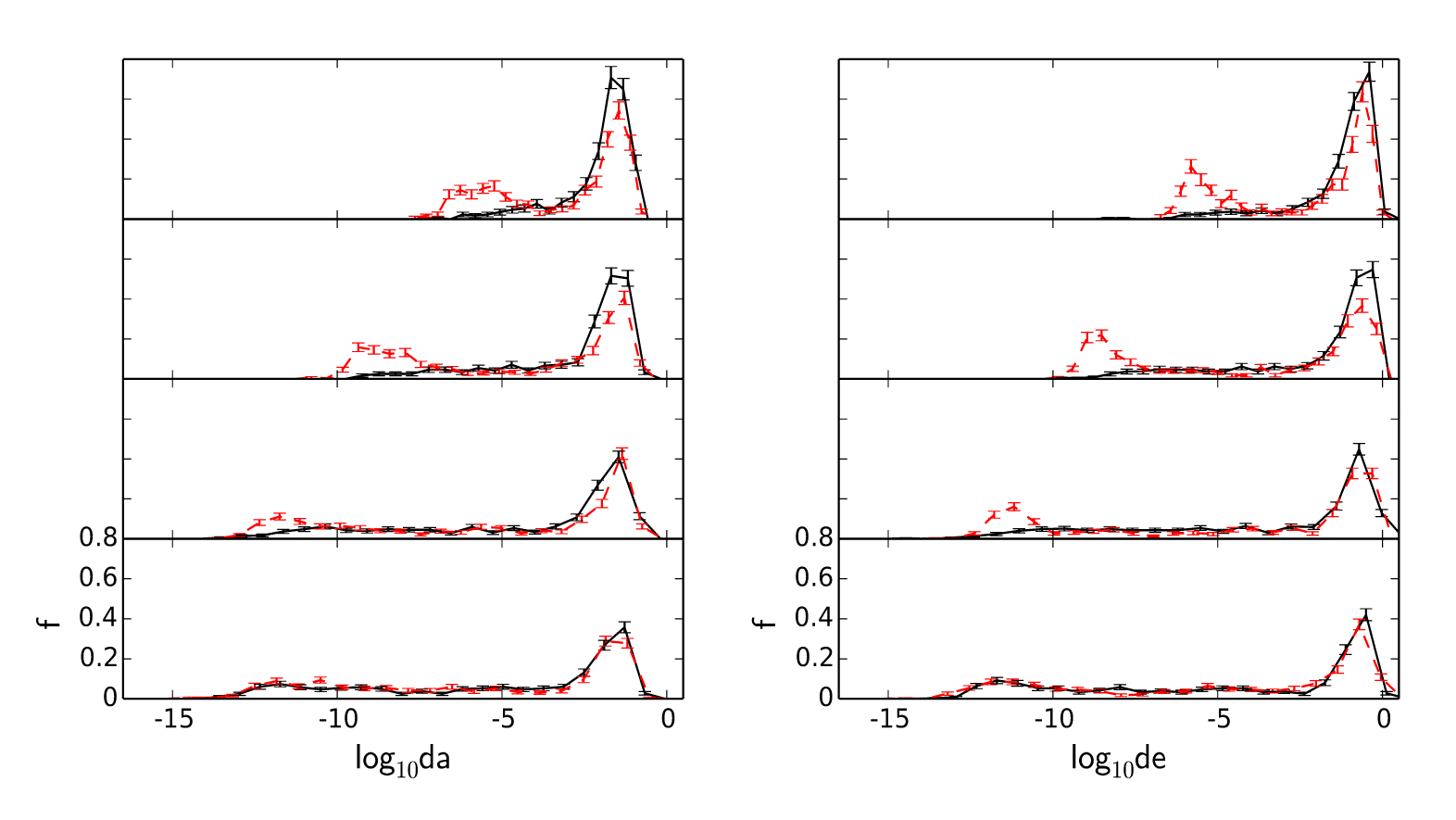} \\
\end{tabular}  
\caption{\csentence{Symmetry of the error distributions.} 
We show distributions of the errors in semimajor axis (left column) and eccentricity (right column) of the binaries formed in the equal-mass Plummer data set.
This is shown separately for the positive errors (solid, black) and negative errors (dashed red), to investigate the symmetry of the error distribution. From the panels at the top to the bottom, the time-step parameter for \texttt{Hermite} varies as $2^{-5}, 2^{-7}, 2^{-9}$ and $2^{-11}$. An asymmetry can be observed at the smallest errors. }
\label{fig:histo_da_de_seq}
\end{figure*}

To measure the symmetry of the error distribution, we count the fraction of positive errors (Fig.~\ref{fig:error}, bottom panels). Again for an $\eta < 10^{-1}$, this fraction converges to 0.5. 
A more detailed comparison is given in Fig.~\ref{fig:histo_da_de_seq}, where we compare distribution functions of positive and negative errors. 
In Sec.~\ref{MoC}, we mentioned that in our experiment we define the \texttt{Brutus} solution to be converged when at least 3 decimal places of every coordinate have converged. To investigate the symmetry up to higher precision, we repeated a subset of 1000 simulations. We did this only for the initial conditions with equal-mass stars picked randomly from a virialised Plummer distribution and this time we obtain solutions converged up to the first 15 decimal places. 

We observe that the majority of errors are larger than $\sim 10^{-3}$ and within the statistical error, the positive and negative errors have a similar distribution. For the smallest errors however, we observe an asymmetry in the sense that there are more negative, small errors. The magnitude of the error where this excess occurs is determined by the precision of the integration. For the smallest $\eta$, the excess is below double-precision and thus not observable anymore (see Sec. \ref{sec:disc_small_errors} for more explanation).  

\subsection{Escaper identity}

\begin{figure*}
\centering
\begin{tabular}{l}
\includegraphics[scale=0.88]{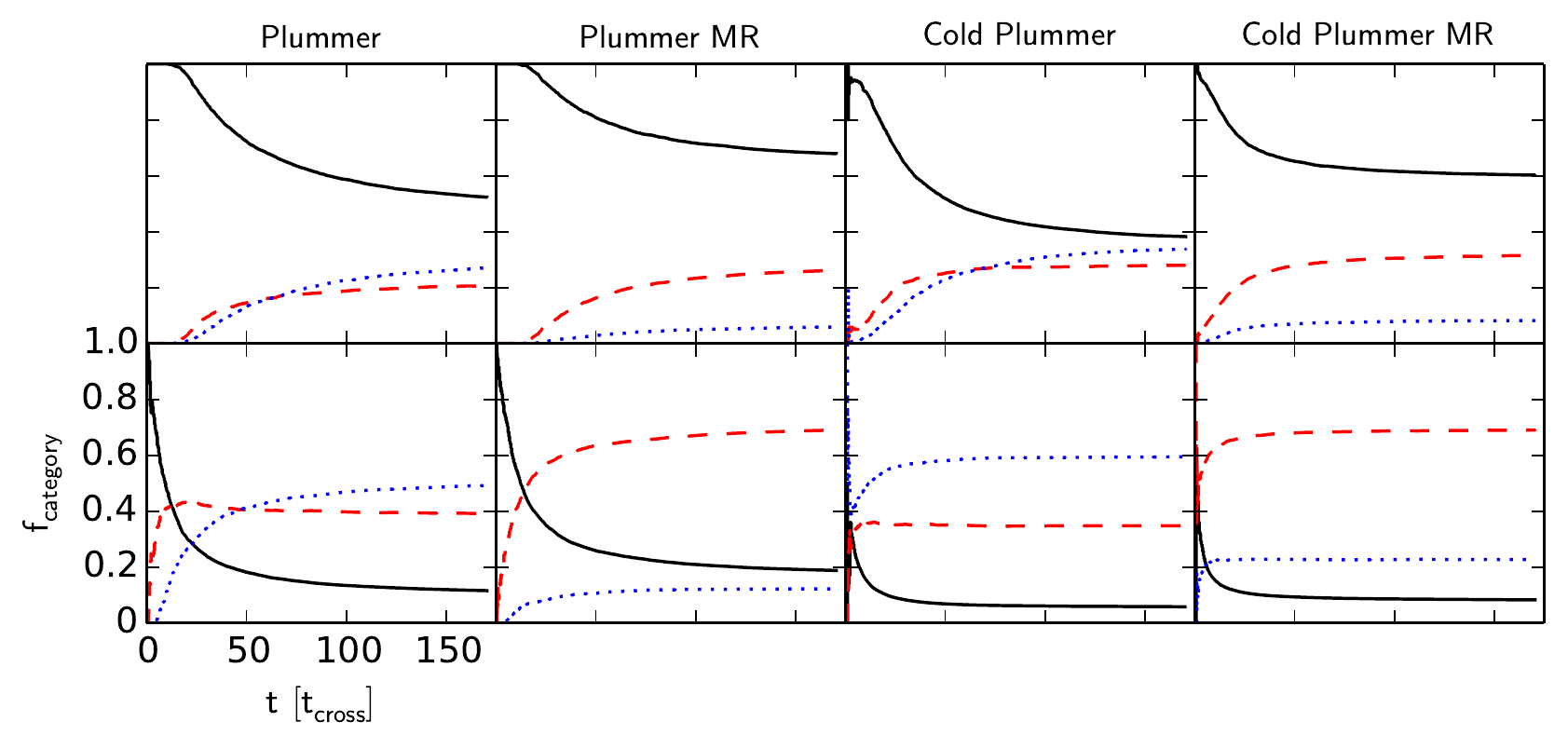} \\
\end{tabular}  
\caption{ \csentence{ The evolution of the relative fractions of categories. } 
The different curves represent the different categories: accurate (solid, black curves), preservation (dashed, red curves) and exchange (dotted, blue curves). These three categories are defined in the text. From left to right, the data are from the Plummer, Plummer with different masses, cold Plummer and cold Plummer with different masses data sets. In the top panels we show the results for a \texttt{Hermite} time-step parameter $\eta=2^{-11}$ and in the bottom for $\eta= 2^{-3}$.  }
\label{fig:category}
\end{figure*}

In this section we compare the solutions obtained with \texttt{Hermite} and \texttt{Brutus} individually, by looking at which star eventually becomes the escaper and which form the binary. We define preservation if the \texttt{Hermite} and the \texttt{Brutus} solution both have the same star as the escaper. We define it as exchange if the escaping star is different. A further distinction can be made in the preservation category, if the \texttt{Hermite} simulation is also accurate. We can typify each \texttt{Hermite} simulation as follows: 

\begin{itemize}
  \item {\bf Accurate}: The coordinates are accurate, up to at least two digits.
  \item {\bf Preservation}: The coordinates are inaccurate, but same star escapes.
  \item {\bf Exchange}: Different star escapes. 
\end{itemize}


In Fig.~\ref{fig:category} we present the fraction of each category as a function of time. 
As expected, systems which dissolve quickly, hardly have time to develop errors and are categorized as accurate simulations. In time however, because errors grow exponentially, the solutions become inaccurate. The fractions of preservation and exchange start to grow. For a small time-step parameter ($\eta = 2^{-11}$, top row in Fig.~\ref{fig:category}), this growth starts after $\sim$ 20 crossing times for the initially virialised systems. For the initially cold systems, the inaccurate fractions already start to grow after a single crossing time. 

The cold collapse with equal-mass stars is the hardest problem to integrate as the accurate fraction is of comparable magnitude as the preservation and exchange fractions.
The accurate fraction generally remains dominant, with a final fraction varying from about 0.4 for the equal-mass cold Plummer to about 0.7 for the Plummer with different masses. For the lesser precision ($\eta = 2^{-3}$, bottom row in the figure), the accurate fractions decrease to below 0.2.  

In the panels in Fig.~\ref{fig:category}, which include the data for the systems with different masses, preservation is more common than exchange. This can be understood, because due to energy equipartition, the lightest particle will be more likely to escape and therefore the identity is more often correct than in the equal mass case. 
For the equal mass case, the fraction of preservation and exchange is comparable, except in the case of the equal-mass cold Plummer with the low precision ($\eta = 2^{-3}$, the bottom row). 
If we regard the identity of the escaping star to be completely random once the solution has become inaccurate, we would expect the fraction of exchange to be twice the fraction of preservation. This is roughly what we observe in the equal mass cold collapse case with low precision. Because of the low precision and the initial close encounter, solutions will diverge very quickly.  In the panel with the higher precision this trend is not observed because the solutions are less randomised. The preservation category includes solutions which slightly differ from the converged solution only in the escape angle of the escaper. Also the long-lived triples are not taken into account here, which will alter these fractions. 

\section{Discussion}\label{Discussion}

\subsection{Energy conservation}

\begin{figure*}
\centering
\begin{tabular}{l}

\includegraphics[scale=0.98]{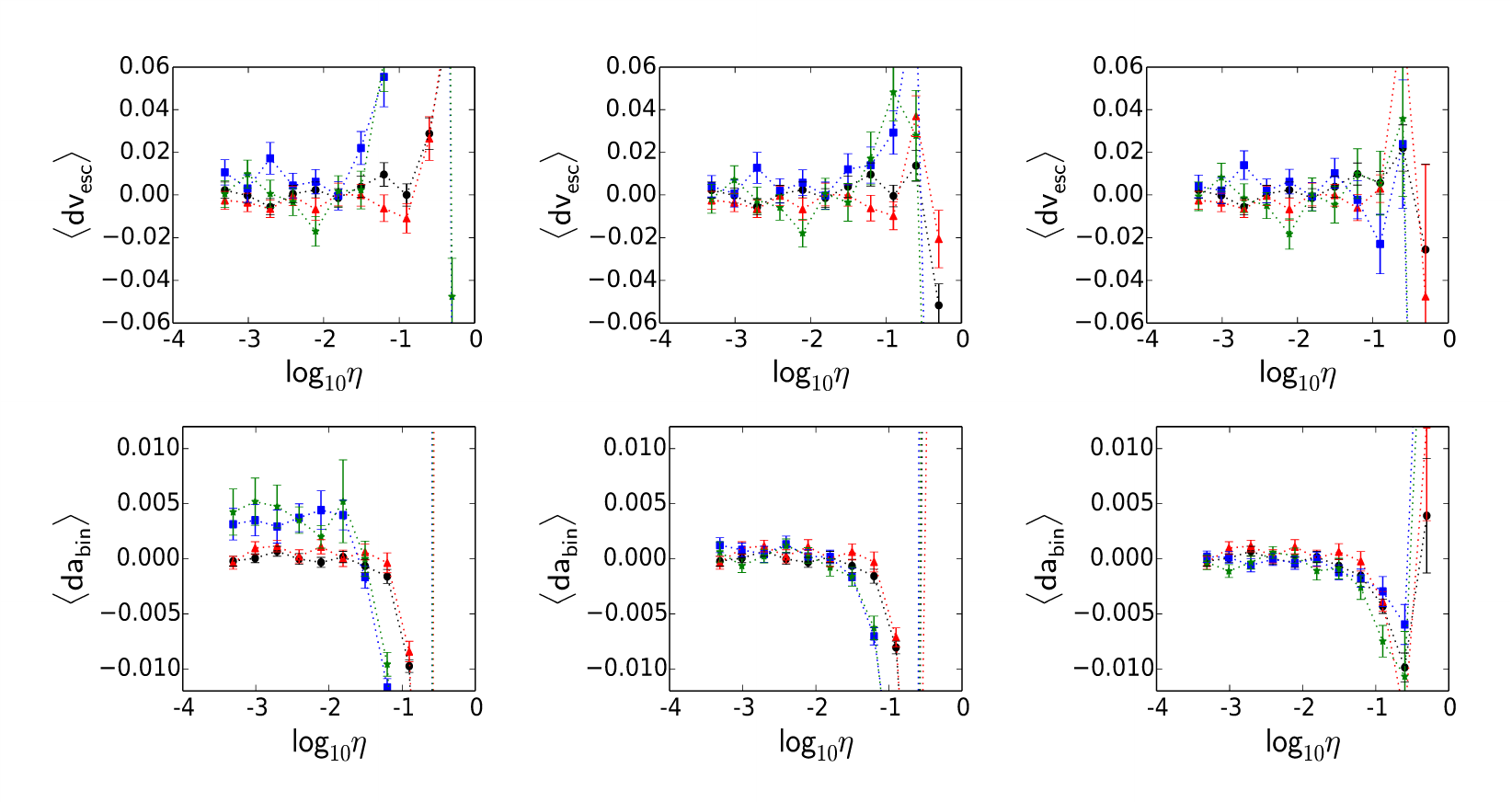} \\

\end{tabular}  
\caption{  \csentence{The effect of cuts in final relative  energy conservation. } 
We plot the average error in the velocity of the escaping star (top row) and the error in the binary semimajor axis (bottom row) as a function of \texttt{Hermite} time-step parameter $\eta$ (with same color coding as in Fig.~\ref{fig:facc}). The three columns differ in the maximum allowed level of relative energy conservation. In the left column we show the results for the total ensemble of solutions, in the middle column for a maximum level of unity and in the right column for 10$^{-1}$. The bias in the left column for the binary semimajor axis is caused by solutions which grossly violate energy conservation. Note that this only happens for the cold collapse simulations. When these outliers are taken out of the ensemble, the bias vanishes. }
\label{fig:dEcut}
\end{figure*}

In every ensemble of \texttt{Hermite} solutions there are some that grossly violate conservation of energy $\left| \Delta E/E \right| > 0.1$. This deformation of the energy hyper-surface in phase-space can allow solutions to reach parts of phase-space which are theoretically forbidden. This affects the global statistical distributions. In Fig.~\ref{fig:dEcut}, we replot the average error in the binary semimajor axis as a function of the time-step parameter. We produce similar diagrams as presented in Fig.~\ref{fig:error}, but this time we  introduce a maximum allowed error in the energy. If we filter out simulations with a relative energy conservation $\left| \Delta E/E \right| > 1$, or $\left| \Delta E/E \right| > 0.1$, we observe that the bias in the average error of the semimajor axis of the binaries vanishes. We conclude that this bias is caused by a few simulations which grossly violate energy conservation. A similar bias in the velocity of the escaping star is less pronounced. 

Time-reversible, symplectic integrators should in principle conserve energy to a better level than non-symplectic integrators, since there is no drift present in the energy error. Therefore, by using a symplectic integrator, the number of simulations with large energy error could be reduced. Using a Leapfrog integrator with constant time-steps, we tested this assumption and we find that for resonant 3-body interactions, it is challenging to obtain accurate solutions. The main reason is that, contrary to regular systems like, for example, the Solar System, resonant 3-body interactions often include very close encounters, which need a very small time-step size to be resolved accurately. This is especially the case for the initially cold systems. Adopting such a small time-step size for the whole simulation, will increase the wall-clock time to that of \texttt{Brutus} or beyond. 

\subsection{Asymmetry at small errors}\label{sec:disc_small_errors}

\begin{figure*}[p]
\centering
\begin{tabular}{l}

\includegraphics[scale=0.98]{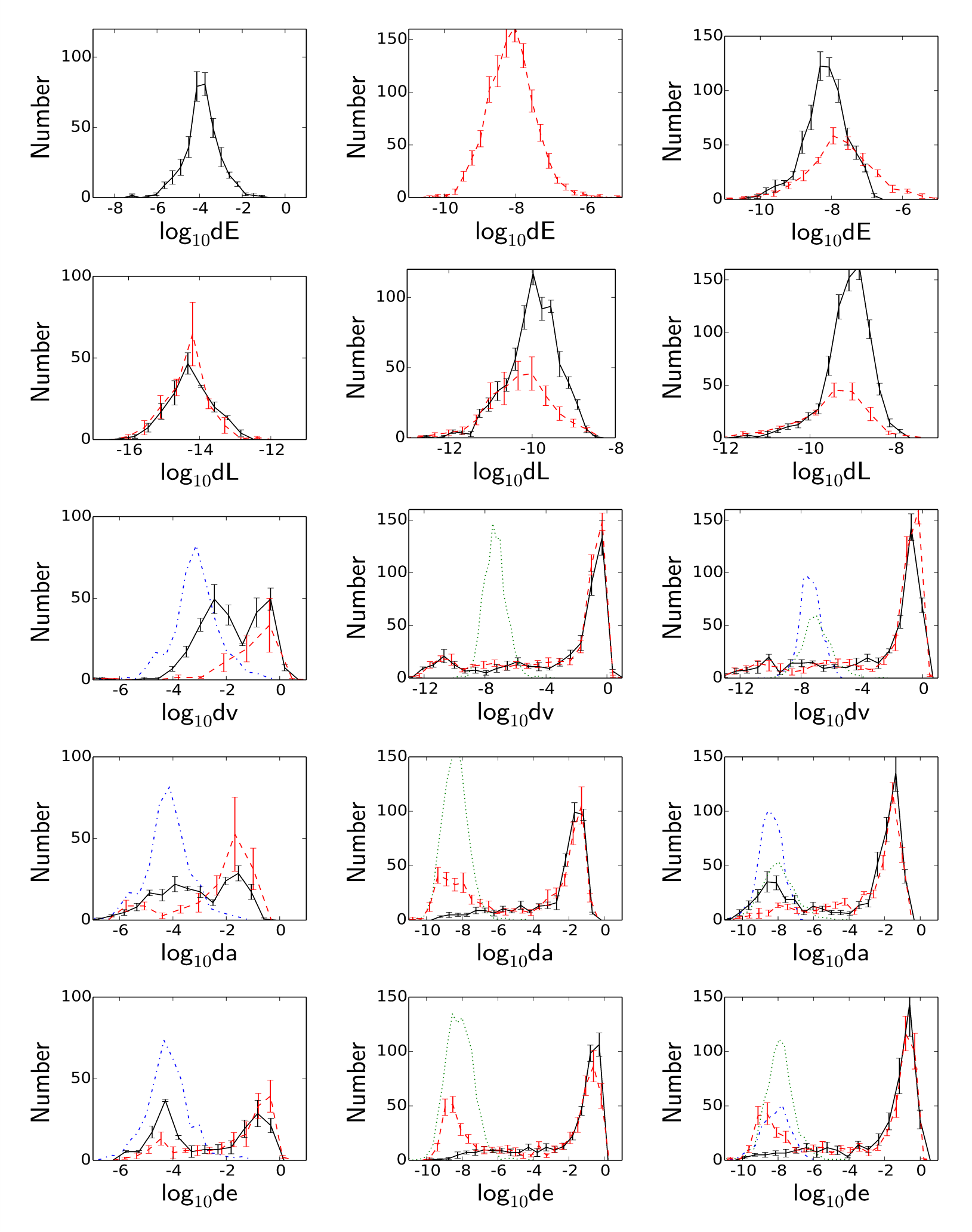} \\

\end{tabular}  
\caption{ \csentence{Explanation of the asymmetry at small errors. } 
We show distributions of the positive (solid, black) and negative (dashed, red) errors in the total energy (top row), total angular momentum (second row), escaper velocity (third row), binary semimajor axis fourth row) and eccentricity (bottom row). 
This is shown for different algorithms: Leapfrog (left column), standard \texttt{Hermite} (middle column) and \texttt{Hermite} with $P(EC)^{n}$ method (right column, $n=3$). Each method implements a shared, adaptive time-step criterion according to Eq.~\ref{eq:1}, with a time-step parameter $\eta = 2^{-7}$. Each of these three integrators has a different asymmetry in the conservation of energy and angular momentum. 
By propagating these asymmetric errors as a small perturbation to the converged solution, we can estimate the resulting asymmetry in the derived quantities. These estimated error distributions are also given separately for the positive (dot-dash, blue) and negative (dotted, green) errors. 
We observe that the estimated error distributions are located at the asymmetry in the empirical error distributions. The asymmetry at small errors is caused by a bias in the integrator. }
\label{histo_dE_dL_he}
\end{figure*}

In Sec.~\ref{symresult}, we discussed an asymmetry at small errors. In Fig.~\ref{histo_dE_dL_he}, we present similar diagrams as in Fig.~\ref{fig:histo_da_de_seq} for the positive and negative errors. This time we add the errors in the total energy and angular momentum of the system and the error in the velocity of the escaper. 

We also vary the integration method because different methods produce different (biased) error distributions in energy and angular momentum. We use a standard Leapfrog integrator, a standard \texttt{Hermite} integrator and a \texttt{Hermite} integrator which uses the P(EC)$^{n}$ method (we adopted n$=$3) \citep{1998MNRAS.297.1067K}. This last method adds an iterative procedure to the algorithm to improve the predictions and corrections, which improves the time-symmetry. For each method we implement a shared, adaptive time-step criterion as in Eq.~\ref{eq:1}, with a time-step parameter $\eta = 2^{-7}$. As a consequence  they will not be time-symmetric nor symplectic. 

We first look at the error distributions in the total energy and angular momentum. We observe that none of them are symmetric, in the sense that the positive and negative errors have identical  distributions, except for the angular momentum in the Leapfrog simulations. The Leapfrog solutions tend to gain energy, whereas the standard \texttt{Hermite} loses energy. The \texttt{Hermite} with the P(EC)$^{n}$ method produces both positive and negative errors in the energy, but not in a symmetric manner.  

To investigate whether the bias in energy and angular momentum conservation propagates to a bias in the binary and escaper properties, we estimate what the errors should be if we regard the error in the energy and angular momentum as a small perturbation to the converged solution.
For the error in the velocity of the escaper, using the derivative of the kinetic energy with respect to velocity, we obtain the following expression:

\begin{equation}
\delta v = \frac{1}{mv} \delta E.
  \label{eq:11}
\end{equation}

\noindent Here $m$ is the mass of a star, $v$ the velocity as obtained by \texttt{Brutus}, $\delta E$ the energy error and $\delta$v the error in the velocity due to this energy error. 
For the binary semimajor axis we obtain: 

\begin{equation}
\delta a = \frac{2}{m^{2}} a^{2} \delta E.
  \label{eq:12}
\end{equation}

\noindent Here $a$ is the semimajor axis from the \texttt{Brutus} solution. 
For the eccentricity we obtain: 

\begin{equation}
\delta e = \frac{1}{\sqrt{1 + \frac{2 \epsilon l^{2}}{\mu^{2}} }} (\frac{l^{2}}{\mu^{2}} \delta \epsilon + \frac{2 \epsilon l}{ \mu^{2} } \delta l).
  \label{eq:13}
\end{equation}

\noindent Here $\mu$ is the total mass of the binary, $\epsilon$ and $l$ the specific energy and specific angular momentum of the binary as obtained by \texttt{Brutus}. The error in the eccentricity $\delta e$ has contributions from errors in the energy $\delta \epsilon$ and angular momentum $\delta l$. 

If we compare the resulting error distributions to the actual error distributions, we find that the approximated error distribution is positioned at the asymmetry in the empirical error distribution. This is most clearly seen for the semimajor axis and eccentricity (see Fig.~\ref{histo_dE_dL_he}).  

The reason why the approximated error distribution overestimates the excess, is because not all errors are solely due to an error in the energy and angular momentum. In time, the numerical solution diverges from the true solution and this error due to divergence will become more dominant. With this in mind, we can approximate the error in a statistic as follows: 

\begin{equation}
\delta S = \delta S_{\mathrm{conservation}} + \delta S_{\mathrm{divergence}}. 
  \label{eq:14}
\end{equation}

\noindent Here $S$ is a statistic that is related to energy and/or angular momentum, $\delta S_{\mathrm{conservation}}$ is the error due to a small perturbation in the energy and/or angular momentum and $\delta S_{\mathrm{divergence}}$ is the error due to divergence of the solution. When the solution has not diverged appreciably yet, the first type of error will dominate and possible biases can be observed. When the second type of error dominates, we observe that the symmetry is restored to within the statistical error. 

Upon inspection of the velocity data, we observe no asymmetry in the \texttt{Hermite} results. When we measure which fraction of the energy error is reserved for the binary and which fraction for the escaper, we find that in most cases the error propagates to the binary. For the Leapfrog however, the asymmetry is still present. 

\subsection{Preservation of the macroscopic properties}\label{MacroProp}


Valtonen et al. \citep{2004ASPC..316...45V} state that the final statistical distributions forget the specific initial conditions
and only depend on globally conserved quantities. This assumption makes predictions which are verified by our experiment. 
The results show that for a time-step parameter $\eta < 2^{-5}$, the distributions are statistically indistinguishable, even though at least half of the solutions diverged from the converged solution. If however, energy conservation 
is grossly violated, biases are introduced in the statistics. In our experiment, a maximum level of relative energy
conservation of $\left| \Delta E/E \right| = 0.1$ was sufficient to remove the biases. This is a much milder constraint than the 
$\left| \Delta E/E \right| \sim 10^{-6}$ usually adopted in collisional simulations. 
Whether 0.1 is also sufficient for systems with more stars, should be  verified experimentally. Heggie \citep{1991pscn.proc...47H} for example, finds that the energy of escaping stars in higher-N systems, depends sensitively on integration accuracy. The maximum required level of energy conservation should be such that it is below the energy taken away from the cluster by the escaping stars. 

The chaoticity of the 3-body problem is illustrated by the scatter diagrams in Fig.~\ref{fig:scatter}. 
For a certain value of a statistic obtained by \texttt{Brutus}, any other value in the allowed outcome space is reachable for the \texttt{Hermite} integrator. For example, if the converged solution gives an eccentricity for the binary of 0.6, a diverged solution can produce any eccentricity between 0 and 1. Once the solution has diverged from the true solution, it will start a random walk through or near the allowed phase-space until the 3-body system has dissolved. We observed that this randomisation happens in such a way that the available outcome space is still completely sampled and that it preserves global statistical distributions. 


In Sec.~\ref{Sect:Experiments}, we discussed that the lifetime of an unstable triple 
does not depend on the integrator used nor on the accuracy of that integrator. 
This last point should be interpreted in the sense that when more effort is put into performing
simulations with higher precision, that this does not change the global statistics,  
even though individual solutions will change with precision (see for example the \texttt{Hermite} results in Fig.~\ref{fig:pyth}). If instead we continue to decrease the precision, there will be a point where biases start to appear. 
Urminsky \citep{2008IAUS..246..235U} analysed the 3-body Sitnikov problem and showed that the precision of the integration influences the average lifetime of triple systems, contrary to our results. The integration times in our experiment however, are much shorter. 
Obtaining a converged solution for a resonant 3-body system for longer than 200 crossing times, is still computationally challenging. Therefore any statistical difference on the long term will not be visible in our experiment.




\section{Conclusion}

\texttt{Brutus} is an \mbox{N-body} code that uses the Bulirsch--Stoer method to control
discretisation errors, and arbitrary-precision arithmetic to control round-off errors. 
By using the method of convergence, where we systematically vary the Bulirsch--Stoer
tolerance parameter and the word-length, we can obtain a solution for a 
particular \mbox{N-body} problem, for which the first $p$ digits in the mantissa are independent of the time-step size and word-length. We call this solution converged to $p$ decimal places. 

Obtaining the converged solution is computationally expensive, mainly because of the exponential divergence of the solution. In some cases, Bulirsch--Stoer tolerances of $10^{-100}$ are needed to reach convergence. We estimate that the time for simulating a star cluster up to core collapse, until convergence, scales approximately exponentially with the number of stars. Simulations with 256 stars however, may be performed within a year of computing time. 

The motivation to obtain expensive, converged solutions is to test the assumption
that the statistics of an ensemble of approximate solutions, are indistinguishable
from the statistics of an ensemble of true solutions. To put this assumption to the test, 
we have investigated the statistics on the breakup of 3-body systems. In our experiment, 
a bound triple system will eventually dissolve into a binary and an escaping star. 
Solutions to every initial realisation were obtained using the standard \texttt{Hermite} integrator and
using \texttt{Brutus}. 

For systems with a long lifetime it is challenging to obtain the
converged solution. Due to repeated ejections and resonances, many accurate digits will be lost and so a very small Bulirsch--Stoer tolerance is required. Therefore, we have set an integration limit at $\sim$ 180 crossing times. 
For equal-mass, virialised systems, $\sim$ 40$\%$ of the random initial realisations were not 
dissolved by this time. For the initially cold systems with different masses this was $\sim$ 10$\%$. 
\texttt{Hermite} and \texttt{Brutus} are consistent on the average lifetime of an unstable triple system. 
However, possible differences on the long term are not visible in this experiment. 

When we compare the results on an individual basis, we find that on average about half of the \texttt{Hermite} solutions
give accurate results, i.e. at most a 1$\%$ relative difference compared to \texttt{Brutus}. 
For the inaccurate results, the error distribution becomes unbiased and symmetric for a time-step parameter $\eta \le 2^{-5}$
and implementing a maximum level of relative energy conservation of $\left| \Delta E/E \right|  < 0.1$.  

Once the conventional solution has diverged from the converged solution, it will start a random walk through or near the allowed region in phase space. such that any allowed outcome of a statistic is reachable. 
This randomisation process completely samples the available outcome space of a statistic and it also preserves the global statistical distributions. 

Kolmogorov--Smirnov tests were performed to compare the global distributions produced by \texttt{Hermite} and \texttt{Brutus}. 
No significant differences were detected when using the criteria mentioned above for the time-step parameter $\eta$ and relative energy conservation. 
This research for the 3-body problem supports the assumption that results from conventional \mbox{N-body} simulations are valid in a statistical sense. We observed however that a bias is introduced for the smallest errors, if the algorithm used to solve the equations of motion, is biased in the conservation of energy and angular momentum. In this research however, this bias did not have an appreciable effect. 
It is important to see whether this remains true for statistics of higher-N systems or systems with a dominant mass. An example of a higher-N system where precision might play a role is a young star cluster (without gas) going through the process of cold collapse \citep{2014MNRAS.445..674C}. 
At the moment of deepest collapse, a fraction of stars will obtain large accelerations, so that a small error in the acceleration can cause large errors in the position and velocity. The rate of divergence can increase up to about 5 digits per Hénon time unit for 128 particles and it increases with N. 


\begin{backmatter}
 
\section*{Competing interests}
  The authors declare that they have no competing interests.

\section*{Author's contributions}
TB wrote the \texttt{Brutus} \mbox{N-body} code, participated in designing the experiments, performed the \mbox{N-body} simulations, gathered the results from the simulations, interpreted the results and wrote the major part of the manuscript. 
SPZ thought of the concept of the \texttt{Brutus} code, participated in designing the experiments, interpreted the results and helped to draft the manuscript. All authors read and approved the final manuscript. 

\section*{Acknowledgements}

We thank Douglas Heggie, Piet Hut, Michiko Fujii and Guilherme Gon\c{c}alves Ferrari for useful discussions and comments on the manuscript. T.B. would also like to thank Ann Young and Lucie J\'{i}kov\'{a} for carefully reading the manuscript and improving the presentation.   
The authors also thank the referees for providing useful improvements to our manuscript. This work was supported by the Netherlands Research Council NWO (grants
\#643.200.503, \#639.073.803 and \#614.061.608) and by the Netherlands
Research School for Astronomy (NOVA).  
Part of the numerical computations were carried out on the Little Green Machine at Leiden
University and on the Lisa cluster at SURFSara in Amsterdam.

\theendnotes


\bibliographystyle{bmc-mathphys} 
\bibliography{brutus}      









\section*{Additional Files}
%
    
\subsection*{Additional file 1 --- Initial and final configurations for the equal-mass Plummer}
This table consists of 10000 initial configurations for three equal-mass stars drawn randomly from a Plummer distribution, together with the final configurations as obtained by \texttt{Brutus}. Additional information is given on the dissolution time, the Bulirsch-Stoer tolerance and word-length. For the configurations which took longer than 500 Hénon time units to dissolve, we give the last configuration of the simulation. For the simulations where the CPU time was very high, we set the final coordinates equal to zero. 
\subsection*{Additional file 2 --- Initial and final configurations for the Plummer with different masses}
Similar as the previous additional file, but for the virialised Plummer initial condition with different masses. 
\subsection*{Additional file 3 --- Initial and final configurations for the cold Plummer}
Similar as the previous additional file, but for the equal-mass Plummer starting with zero velocities. 
\subsection*{Additional file 4 --- Initial and final configurations for the cold Plummer with different masses}
Similar as the previous additional file, but for the Plummer with different masses, starting with zero velocities.

\end{backmatter}
\end{document}